\documentclass[prb,twocolumn,showpacs,superscriptaddress,floatfix]{revtex4}

\usepackage{amsmath}
\usepackage{amstext}
\usepackage{amssymb}
\usepackage{latexsym}
\usepackage[dvips]{graphicx}
\usepackage{bbm}

\newcommand{\beq}{\begin{equation}}
\newcommand{\eeq}{\end{equation}}
\newcommand{\barr}{\begin{eqnarray}}
\newcommand{\earr}{\end{eqnarray}}
\newcommand{\ket}[1]{\left\vert#1\right\rangle}
\newcommand{\bra}[1]{\left\langle#1\right\vert}
\newcommand{\Ham}{\mathcal H}
\newcommand{\eps}{\varepsilon}
\newcommand{\tr}{\,{\rm tr}\,}
\newcommand{\mm}[1]{{\mathbf{#1}}}
\def\vec#1{\underline{#1}}
\def\ave#1{\langle #1\rangle}

\begin{document}

\title{Charge and spin transport in strongly correlated
  one-dimensional quantum systems driven far from equilibrium}

\author{Giuliano Benenti}
\affiliation{CNISM, CNR-INFM, and Center for Nonlinear and Complex Systems,
  Universit\`a degli Studi dell'Insubria, Via Valleggio 11, 22100
  Como, Italy}
\affiliation{Istituto Nazionale di Fisica Nucleare, Sezione di Milano,
  Via Celoria 16, 20133 Milano, Italy}

\author{Giulio Casati}
\affiliation{CNISM, CNR-INFM, and Center for Nonlinear and Complex Systems,
  Universit\`a degli Studi dell'Insubria, Via Valleggio 11, 22100
  Como, Italy}
\affiliation{Istituto Nazionale di Fisica Nucleare, Sezione di Milano,
  Via Celoria 16, 20133 Milano, Italy}
\affiliation{Centre for Quantum Technologies,
  National University of Singapore, 117543 Singapore }

\author{Toma\v z Prosen}
\affiliation{Physics Department, Faculty of Mathematics and Physics,
  University of Ljubljana, SI-1000 Ljubljana, Slovenia}

\author{Davide Rossini}
\affiliation{International School for Advanced Studies (SISSA),
  Via Beirut 2-4, I-34014 Trieste, Italy}

\author{Marko \v Znidari\v c}
\affiliation{Physics Department, Faculty of Mathematics and Physics,
  University of Ljubljana, Ljubljana, Slovenia}

\date{\today}

\begin{abstract}
  We study the charge conductivity in one-dimensional
  prototype models of interacting particles, such as the Hubbard
  and the $t-V$ spinless fermion model, 
  when coupled to some external baths injecting
  and extracting particles at the boundaries.
  We show that, if these systems are driven far from equilibrium, 
  a negative differential conductivity regime can arise.
  The above electronic models can be mapped 
  into Heisenberg-like spin ladders coupled to two magnetic baths,
  so that charge transport mechanisms are explained in terms
  of quantum spin transport.
  The negative differential conductivity is due to oppositely
  polarized ferromagnetic domains which arise at the edges of the chain,
  and therefore inhibit spin transport: we propose a qualitative understanding
  of the phenomenon 
  by analyzing the localization of one-magnon excitations created
  at the borders of a ferromagnetic region.
  We also show that negative differential conductivity 
  is stable against breaking of integrability.
  Numerical simulations of non-equilibrium time evolution have been performed
  by employing a Monte-Carlo wave function approach
  and a matrix product operator formalism.
\end{abstract}



\maketitle

\section{Introduction}

Transport properties of strongly interacting fermions in microscopic
models of one dimensional quantum systems have been the subject of a large
number of theoretical and experimental studies~\cite{baeriswyl:BOOK}.
In the last few years this has become a topical subject,
due to the rapidly developing process of miniaturization in semiconductor
microelectronics devices that is approaching its natural limits, reaching
the atomic or molecular scale~\cite{joachim00,nitzan,richter}.
Of course, a technological breakthrough in this direction would require
conceptually new devices, such as few or even single molecules
embedded between electrodes, that could perform the basic functions
of microelectronics.
The first promising step in the realization of such devices comes
from the observation of many-body effects, as the Coulomb blockade
and the Kondo effect in nanometer-scale systems, like single molecules
or carbon nanotubes~\cite{park02,nygard00}.
Establishing and reaching a suitable degree of control
of nonlinear electronic transport, such as a 
{\em Negative Differential Conductivity} (NDC) regime, would be one
of the ultimate tasks for functional nanodevices, since it lays
at the basis of current rectification and amplification.
NDC has been observed in a variety of nanoscopic objects, like
semiconductor quantum dots~\cite{weis93}, carbon nanotubes~\cite{zhou00},
as well as single molecules~\cite{chen99}.

From a theoretical point of view, nonlinear transport properties in
such systems are usually studied by considering effective models
of few single-particle levels, see
e.g. Refs.~\onlinecite{averin91,beenakker91,weinmann95,thielmann05,elste06}.
In this paper we adopt a rather different perspective and
study the full many-body quantum dynamics of one-dimensional prototype models
of strongly interacting fermions,
when they are coupled to some external baths.
We will show how effects of nonlinear transport naturally emerge
in far-from-equilibrium situations, by exploiting
the many-body dynamics of such microscopic models in its whole complexity.
While situations close to equilibrium are quite well understood
and can be tackled by the powerful linear response
formalism~\cite{loss03,zotos99,prelovsek04,meisner03,narozhny98,castella95,zotos96},
almost nothing is known about the physics of such systems far from equilibrium.
In this regime new quantum phases and phenomena can appear, thus making
the problem relevant also for fundamental physics~\cite{prosen08b}.
Furthermore, the study of far-from-equilibrium quantum systems 
is of interest also for issues such as the control of heat flow at the
nanoscale~\cite{mahler,baouwen08} and, in quantum information processing,
for quantum state preparation/transfer~\cite{vittorio}.
Unfortunately a fully analytical treatment is generally unfeasible~\cite{prosen08},
and one typically has to resort to numerical simulations, aimed at solving
the quantum master equation~\cite{saito00,saito03,michel03,breuer,monasterio,paper1,prosen},
or based on different approaches, like path integral Monte Carlo approach \cite{rabani08},
time-dependent density matrix renormalization
group or current density functional theory~\cite{langer08,polini08,xianlong08},

In this paper we consider two prototype microscopic one-dimensional models
of interacting fermions, namely the Hubbard model and
the $t-V$ spinless fermion model, and couple them to some external baths
that inject and extract particles at the system edges, thus mimicking
the effect of electrodes.
The Hubbard and the $t-V$ model can be mapped into the Heisenberg spin-$1/2$
ladder and chain, respectively. In these spin models NDC
reflects in the suppression of spin conduction, while the operators
injecting/extracting electrons are mapped into operators
flipping the two spin species at the border of the chain.
As an example of spin chains coupled to such ``magnetic baths''
one can consider
molecular spin wires~\cite{bogani08,ghirri07} with each boundary
coupled to an external spin (magnetic impurity); the ratio
of up/down and down/up spin-flip probabilities is determined 
by the populations of such impurities, which in turn can be 
tuned by means of applied electromagnetic fields.
In the linear response regime, the electronic (i.e., fermionic) transport
and correspondingly the spin transport can be ballistic or diffusive,
depending on the values of the Hamiltonian parameters.
Here we focus on the {\it far-from-equilibrium regime}, beyond the
linear response regime. Our numerical results 
show that, strikingly, in the above 
mentioned models it is possible to achieve a regime
where charge/spin conductivity exhibits a negative differential with 
respect to the driving strength.
NDC arises as a result of the appearance of a far-from-equilibrium 
steady state characterized, for the spin chain models, by 
long-range spin ordering into ferromagnetic domains.
These ferromagnetic domains correspond to charge separation
in the fermionic models, with all the electrons frozen in half of the 
lattice. In both cases, it is clear that such cooperative many-body
state hampers spin flips (or charge injection/extraction), thus 
strongly suppressing the current. 
We will show that our numerical results can be qualitatively explained in 
terms of localization of one-magnon excitations.

The paper is organized as follows: in Sec.~\ref{sec:Methods} we start
by setting our electron transport problem and reducing it to a
Lindblad master equation formalism, that will be used
throughout the paper.
In Sec.~\ref{sec:Hubbard} we introduce the model of open Hubbard chain coupled
to two macroscopic reservoirs, and discuss some peculiar charge transport
properties, focusing on the NDC behavior.
In Sec.~\ref{sec:tVmodel} we consider a simplified model for spinless
fermions and show that it can be mapped into a Heisenberg spin chain.
In Sec.~\ref{sec:Heisenberg} 
we study in details the spin transport
properties of the Heisenberg chain. 
Moreover, we provide a one-magnon localization 
argument which qualitatively explains the observed NDC behavior.
To explore the possibility that our system undergoes  
a metal-insulator phase transition
when driven far from equilibrium, we propose to study
steady-state spin-spin correlation functions.
We also add, in Sec.~\ref{sec:stagger}, 
a staggered magnetic field and check that NDC is stable
against breaking of integrability.
Finally, in Sec.~\ref{sec:conclusions} we draw our conclusions.
In the Appendices we describe the two numerical methods 
used throughout the paper,
namely the quantum trajectories approach and the matrix product operator
formalism (App.~\ref{app:numerical}),
give technical details on the mapping of
our fermionic systems into spin chain models (App.~\ref{app:mapping}),
provide some numerical results about the
steady-state spin-spin correlation functions (App.~\ref{app:spinspin}),
and present an analytical derivation of the one-magnon argument for
the Heisenberg spin chain (App.~\ref{app:magloc}).
A brief account of the NDC features of the Heisenberg chain
can be found in a recent paper by some of us~\cite{paper1}.

\section{Master equation approach} \label{sec:Methods}

Our electronic transport model is described by the Hamiltonian
\begin{equation}
\Ham = \Ham_S + \Ham_{l} + \Ham_c,
\end{equation}
where the different terms correspond to the nanoscale electronic
system, the leads, and the leads-system coupling, respectively. 

\begin{figure}[!t]
  \begin{center}
    \includegraphics[scale=0.40]{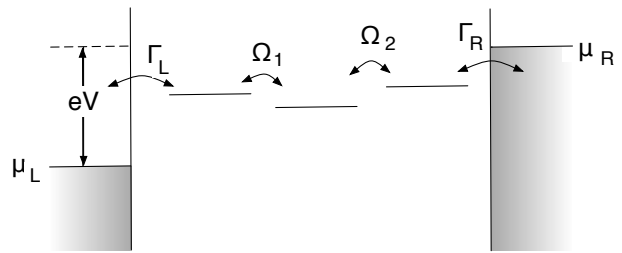}
    \caption{Schematic drawing of the level structure for
    a chain with $N=3$ sites. We assume that the lead-chain tunneling 
    rates $\Gamma_L,\Gamma_R$ are much larger than the intra-chain 
    tunneling rates $\Omega_1,\Omega_2$. The electronic current flows from 
    the right lead (emitter) to the left lead (collector).}
    \label{fig:wire}
  \end{center}
\end{figure}

As sketched in Fig.~\ref{fig:wire},
we consider a $N$-site chain, whose autonomous dynamics is 
described by Hamiltonian $\Ham_S$. Such a lattice models
a nanoscale system, for instance a chain of coupled quantum dots 
or a molecular wire (in the latter case, each lattice site corresponds
to one atom). 

The first and the last site of the chain, $1$ and $N$, are coupled 
to the left/right leads via the tunneling Hamiltonian
\begin{equation}
\Ham_c=\sum_{k,s} (T_{Lk} c_{Lk,s}^\dagger c_{1,s} +
T_{Rk} c_{Rk,s}^\dagger c_{N,s}) + {\rm H.c.},
\end{equation}
where $c^\dagger, c$ are fermionic creation/annihilation operators:
$c_{js}^\dagger$ creates an electron with spin $s$ at site $j$
($j=1,...,N$, $s=\uparrow,\downarrow$), 
$c_{Lk,s}^\dagger$ ($c_{Rk,s}^\dagger$) creates an 
electron in the left (right) lead in the state $|Lk,s\rangle$
($|Rk,s\rangle$). 

The leads are modeled as ideal Fermi gases, 
\begin{equation}
\Ham_{l}=\sum_{k,s}
\epsilon_{k} (c_{Lk,s}^\dagger c_{Lk,s}+c_{Rk,s}^\dagger c_{Rk,s}),
\end{equation}
which are initially at equilibrium, at temperature ${\mathcal T}$
and chemical potentials $\mu_L$, $\mu_R=\mu_L +e V$,
where $V$ is the applied bias voltage and $e$ the electron charge. 
We assume that the coupling between the system and the leads is weak,
such that the state $\rho_B(t)$ of the leads at any time 
$t$ is well described by 
$\rho_B(t)=\rho_L\otimes\rho_R$, with $\rho_L$ and $\rho_R$
grand canonical density matrices for the left and right leads,
respectively.

A Lindblad master equation for the system's evolution 
can be obtained from our microscopic Hamiltonian model 
following standard textbook derivations, under the usual 
Born-Markov and rotating wave approximations 
and neglecting the Lamb-type renormalization of the
unperturbed energy levels (see, e.g., 
Sec.~3.3 of Ref.~\onlinecite{breuer:BOOK}): 
\beq
   \frac{\partial \rho}{\partial t} = -\frac{i}{\hbar} [ \Ham_S, \rho]
   - \frac{1}{2} \sum_m \{ L_m^\dagger L_m , \rho \}
   + \sum_m L_m \rho L^\dagger_m \, ,
   \label{eq:MasterEq}
\eeq
where $\rho(t)$ is the density matrix describing the open
quantum system,  
the Lindblad operators $L_m$ describe the effect of the environment,
while $[\cdot,\cdot]$ and $\{ \cdot, \cdot \}$
denote the commutator and the anti-commutator, respectively.
Hereafter we shall set $\hbar = 1$.
Moreover, we assume 
that the tunneling between sites $1$ ($N$) and the left (right) 
lead is much faster than the intra-chain tunneling and that 
we can neglect the effects of Coulomb repulsion on the 
system-leads transition rates. This is the case in the so-called
wide-band limit, in which the conduction bandwidth of the leads
is much larger than all other relevant energy scales and all the relevant
lead states are located in the center of the conduction band, so that the
energy dependence in the system-leads transition rate may be neglected.
Under these approximations, we can easily derive the 
Lindblad master equation (\ref{eq:MasterEq}), with  
four Lindblad operators on each of the two chain ends:
\beq
\begin{array}{ll}
   L_1 = \sqrt{\Gamma_L f_{L}}     \, c^\dagger_{1,\uparrow}\,, \quad&
   L_2 = \sqrt{\Gamma_L (1-f_{L})} \, c_{1,\uparrow}\,, \quad \\
   L_3 = \sqrt{\Gamma_L f_{L}} \, c^\dagger_{1,\downarrow}\,, \quad&
   L_4 = \sqrt{\Gamma_L (1-f_{L})} \, c_{1,\downarrow} \, ,
\end{array}
\label{eq:Lindblad}
\eeq
and similarly
\beq
\begin{array}{ll}
   L_5 = \sqrt{\Gamma_R f_{R}}     \, c^\dagger_{N,\uparrow}\,, \quad&
   L_6 = \sqrt{\Gamma_R (1-f_{R})} \, c_{N,\uparrow}\,, \quad \\
   L_7 = \sqrt{\Gamma_R f_{R}} \, c^\dagger_{N,\downarrow}\,, \quad&
   L_8 = \sqrt{\Gamma_R (1-f_{R})} \, c_{N,\downarrow} \, ,
\end{array}
\label{eq:LindbladRight}
\end{equation}
where 
\begin{equation}
\Gamma_L \equiv 2\pi |T_L(E_1)|^2 g_L(E_1),
\quad
f_L \equiv f_L(E_1),
\end{equation}
\begin{equation}
\Gamma_R \equiv 2\pi |T_R(E_N)|^2 g_R(E_N),
\quad
f_R \equiv f_R(E_N),
\end{equation}
with $E_1$ ($E_N$) being the energy difference between the two chain states
involved in the transitions for site $1$ ($N$),
$g_l$ ($l=L,R$) being the density of states of lead $l$ 
(we assume that the leads  are macroscopic objects, with a continuous density of
states), $T_l(\epsilon=\epsilon_k)\equiv T_{lk}$,
and $f_l(\epsilon)=\left[1+e^{(\epsilon-\mu_l)/k_B {\mathcal T}}\right]^{-1}$
denoting the Fermi function, with $k_B$ the Boltzmann constant.
Note that the energy differences $E_1,E_N$ contain the charging 
energy $E_c$, if an electron is tunneling onto an already occupied site,
but does not contain it if the site is initially empty. We have neglected
the dependence of the Fermi functions $f_L,f_R$ on $E_c$. This condition
is fulfilled when $E_c\ll k_B {\mathcal T}$ (in the Hubbard and $t-V$ models 
described in this paper such constraint corresponds to on-site 
repulsion $U\ll k_B {\mathcal T}$ and nearest-neighbor
repulsion $V\ll k_B {\mathcal T}$, respectively). 
Finally, in order to consider incoherent tunneling of electrons
into the chain (sequential tunneling approximation), the level 
broadening due to the chain-leads tunneling must be small compared
to temperature, that is, we require
$\Gamma_L,\Gamma_R\ll k_B {\mathcal T}$.

As we shall see in this paper, a main advantage of the master equation 
approach is that it can be applied far-from-equilibrium,
beyond linear response regime.
The far-from-equilibrium regime in our model
corresponds to large bias voltage,
$eV\gg k_B {\mathcal T}$, with the energy differences $E_1,E_N$
such that $\mu_L\ll E_1,E_N \ll \mu_R$. In this limit, 
$f_L\to 0$, $f_R\to 1$, that is, the backward flow of 
electrons (against the applied bias) vanishes. 

The master equation approach may be generalized, including the 
effects of Coulomb repulsion~\cite{gurvitz} (thus describing the
Coulomb blockade phenomenon) or the coupling of multilevel
nanoscale systems to external leads~\cite{li,mukamel}.
The price to pay for such generalizations is, in general, 
the introduction of a larger number of Lindblad operators, 
corresponding to all possible transitions between the levels 
of the nanosystem.

The main transport quantity, the electron current $j$, is defined 
by the continuity equation of the local charge density 
$n_{k,s}\equiv c_{k,s}^\dagger c_{k,s}$, $n_k = n_{k,\uparrow} + n_{k,\downarrow}$:
\begin{equation}
\frac{\partial n_{k}}{\partial t} + \nabla j_k=0,
\end{equation}
which can be rewritten as 
\begin{equation}
j_{k+1}-j_k={i} [n_{k},\Ham_S], \quad k=1,\ldots ,N-1.
\label{eq:current1}
\end{equation}
Note that, due to the continuity equation, one has 
$j=j_k$ for any $k$ along the chain.

By definition the electron current is given by 
\begin{equation}
\langle j \rangle =\frac{d N_R}{dt}=-\frac{d N_L}{d t},
\label{eq:current2}
\end{equation}
with $N_R$ ($N_L$) being the number of electrons in the right (left) lead.
This equation
expresses the current in terms of the number of electrons which enter
the system from the left reservoir ($-d N_L/dt$),
or go out into the right reservoir ($d N_R/dt$) per time unit.
As we shall discuss in Appendix~\ref{app:numerical}, 
$-d N_L/dt$ and $d N_R/dt$ may be computed by means of 
the quantum trajectories approach.
We will use both Eqs.~(\ref{eq:current1}) and 
Eq.~(\ref{eq:current2}) to compute the current.

Explicitly solvable models of master equations 
are very limited, therefore support
from extensive numerical simulations is generally needed.
We used two methods to face this problem.
The first is a Monte-Carlo wave function approach, that is based on the
technique of Quantum Trajectories (QT), widely used in quantum
optics~\cite{dalibard92,plenio,brun,carlo04}.
The second is a Matrix Product Operator (MPO) technique based on the
time-dependent density matrix renormalization group
method~\cite{vidal,feiguin,verstraete,zwolak,schollwock}.
QT revealed themselves a powerful tool in the study
of relatively small system sizes, especially for situations
with a strong external bias, where equilibration times needed to reach
the steady state are generally long.
On the other hand, the MPO method can deal with systems up to one order
of magnitude larger, but it may encounter some difficulties in
converging to the stationary state for large driving fields.
Both numerical methods are 
briefly discussed in Appendix~\ref{app:numerical}.

\section{Hubbard model} \label{sec:Hubbard}

We start our analysis by considering a paradigmatic model for the
physics of strongly interacting electronic systems: the Hubbard model.
Its Hamiltonian is a sum of a kinetic term allowing
for electron tunneling between the neighboring lattice sites,
and a potential term consisting of an on-site interaction;
in one dimension it is given by:
\beq
   \Ham_S = -t \sum_{j, s} \left( c^\dagger_{j,s} c_{j+1,s} + {\rm H.c.} \right)
   + U \sum_{j=1}^N n_{j, \uparrow} n_{j, \downarrow} \, ,
   \label{eq:Hubbard}
\eeq
where $s$ stands for spin up $\uparrow$ or down $\downarrow$ configuration,
while $j=1,\ldots, N$ is the site index and $N$ is the number of lattice sites.
The operators $c^\dagger_{j,s} , \, c_{j,s}$ create/annihilate a spin-$1/2$
fermion with spin $s$ at site $j$, and satisfy the usual anticommutation rules;
$n_{j,s} = c^\dagger_{j,s} c_{j,s}$ is the corresponding number operator.
We consider open boundary conditions, therefore
the sum over $j$ in the first term runs from $1$ to $N-1$.
The system parameters $t$ and $U$ ($U>0$) describe, 
respectively, the nearest neighbor
hopping strength and the on-site repulsion between electrons with 
opposite spins.

Both ends of the Hubbard chain are coupled to some electrodes which act on the
system by injecting or extracting particles with different spins.
In the Lindblad master equation formalism, we assume that their effect 
can be modeled by the Lindblad operators 
(\ref{eq:Lindblad}) and (\ref{eq:LindbladRight}).

From the continuity equation (\ref{eq:current1}) we obtain
\begin{equation}
j=-t\sum_s (i c_{k,s}^\dagger c_{k+1,s}+{\rm H.c.}), \quad (k=1,...,N-1).
\label{eq:current3}
\end{equation}

We examined the fermionic transport properties of the Hubbard
model~\eqref{eq:Hubbard} coupled to external baths by exploiting a mapping
of this system into a spin ladder model, where the particle current is replaced
by the spin current.
Specifically, the Hamiltonian in Eq.~\eqref{eq:Hubbard} is mapped into
a Heisenberg spin ladder by first employing a double Jordan-Wigner
transformation of spin-up and spin-down fermions (separately) into
two different species of hard core bosons.
Then they are transformed into two species of spin-$1/2$ particles,
which are described by the Pauli matrices $\sigma_j^\alpha$ and $\tau_j^\alpha$
($\alpha = x,y,z$). Details are given in Appendix~\ref{app:mapping}.
One finally arrives at the following spin ladder Hamiltonian for the 
autonomous system:
\barr
   \nonumber \Ham_S & = & -\frac{t}{2} \sum_{j=1}^{N-1} \Big[
   \big( \sigma^x_j \sigma^x_{j+1} + \sigma^y_j \sigma^y_{j+1} \big) +
   \big( \tau^x_j \tau^x_{j+1} + \tau^y_j \tau^y_{j+1} \big) \Big] \\
   {} & {} & + \frac{U}{4} \sum_{j=1}^N \left( \sigma^z_j + 1 \right)
   \left( \tau^z_j + 1 \right) \, .
   \label{eq:HeisLadder}
\earr
The Lindblad operators in Eqs.~\eqref{eq:Lindblad},~\eqref{eq:LindbladRight}
correspond, in the spin-$1/2$ picture, to operators flipping the two spin
species at the borders of the chain.
Apart from some phase factor which is uninfluent for our purposes
(see Appendix~\ref{app:mapping}) 
we have that 
$c^\dagger_{j,\uparrow} \to \sigma^+_j$ and
$c_{j, \uparrow} \to \sigma^-_j$ for spin-up particles, while
$c^\dagger_{j,\downarrow} \to \tau^+_j$ and
$c_{j, \downarrow} \to \tau^-_j$ for spin-down particles
[$\sigma^\pm_j \equiv (\sigma^x_j \pm {\rm i} \, \sigma^y_j)/2$ and
$\tau^\pm_j \equiv (\tau^x_j \pm {\rm i} \, \tau^y_j)/2$ 
denote the raising/lowering operators for the two spin species].

The spin current $j$ analogous to the electron current (\ref{eq:current3}) 
is derived from the continuity equation
for the local spin operators $S^z_k \equiv \sigma^z_k / 2$:
$\partial_t S^z_k + \nabla (j_\sigma)_k = 0$, which can be rewritten as
$(j_\sigma)_{k+1} - (j_\sigma)_k = \frac{i}{2} [ \sigma^z_k, \Ham_S ]$
(analogous equations can be written for the $\tau$ species 
in Eq.~\eqref{eq:HeisLadder}).
We obtain
\begin{eqnarray}
j &\equiv& j_\sigma + j_\tau,  \label{eq:current4} \\
j_\sigma &=& -\frac{t}{2}(\sigma^x_k \sigma^y_{k+1} - \sigma^y_k \sigma^x_{k+1}), \nonumber \\
j_\tau &=& -\frac{t}{2}(\tau^x_k \tau^y_{k+1} - \tau^y_k \tau^x_{k+1}). \nonumber
\end{eqnarray}

\begin{figure}[!t]
  \begin{center}
    \includegraphics[scale=0.33]{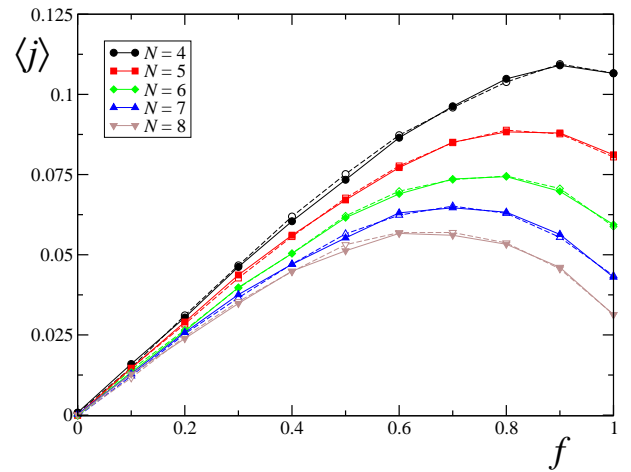}
    \caption{(Color online). Spin current for the $\sigma$ (full curves and
      symbols) and the $\tau$ species (dashed curves and empty symbols)
      of spin as a function of the driving strength for the Hamiltonian
      in Eq.~\eqref{eq:HeisLadder} with $U = 5$. The system-bath
      coupling is set equal to $\Gamma = 0.5$; the simulation time
      (QT approach)
      is $T = 2 \times 10^5$. Note that curves and symbols for the 
      $\sigma$ and $\tau$ species are nearly superimposed.
      For the Hubbard model we set $t=1$ as the system's energy scale.}
    \label{fig:Hubbard_NDC}
  \end{center}
\end{figure}

In the following, we choose a symmetric driving: 
$\Gamma_L=\Gamma_R \equiv \Gamma$ and 
$f_{L,R}
= \frac{1}{2} (1 \mp f)$,
so that $f \equiv f_R - f_L \in [0,1]$ 
($f_L \le f_R$, $0\le f_L, f_R \le 1$)
is the parameter controlling
the driving strength. 
Small $f$ implies that the system is weakly driven by the external baths,
and behaves as in the linear response regime.
In the opposite limiting case $f=1$, the left (right) bath only induces
up-down (down-up) spin flips for both spin species.

Using the method of quantum trajectories we evaluated the stationary
spin currents $\langle j_{\sigma} \rangle$, $\langle j_{\tau} \rangle$
for the two species of spins.
Due to the mapping between the electrons described by the Hamiltonian
in Eq.~\eqref{eq:Hubbard} and the spins obeying Eq.~\eqref{eq:HeisLadder},
this spin current exactly equals the electronic current in the Hubbard model.
In particular, $\langle j_{\sigma } \rangle$ ($\langle j_{ \tau} \rangle$)
is the current flow of electrons with spins pointing up (down), that is
the crucial physical quantity in charge transport.

Perhaps the most interesting result we found in the current behavior as a
function of the driving is the emergence of a NDC phenomenon for sufficiently
strong drivings, as shown in Fig.~\ref{fig:Hubbard_NDC}.
It happens that, while for small $f$ values the current increases,
there exists a value $f^*$ at which $\langle j \rangle$ exhibits a maximum
and then, further increasing $f$, it decreases.

\begin{figure}[!t]
  \begin{center}
    \includegraphics[scale=0.33]{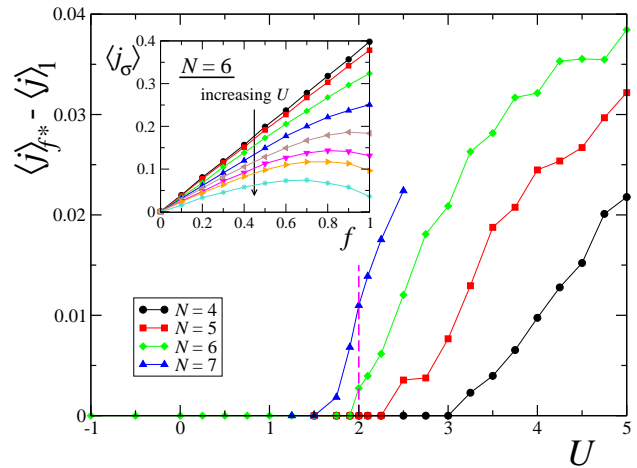}
    \caption{(Color online). Maximum current minus current at maximum
      driving strength, $\langle j \rangle_{f^*} - \langle j \rangle_1$
      as a function of the on-site repulsion $U$.
      In the inset we show the spin current as a function of $f$, for
      a fixed size $N=6$ and different values of $U$:
      from top to bottom $U=$ 0 (circles), 0.5 (squares), 1 (diamonds),
      1.5 (triangles up), 2 (triangles left), 2.5 (triangles down),
      3 (triangles right), 5 (stars).
      Data are for $\Gamma = 1$, and the simulation time 
      (QT approach) is $T=10^5$.
      The current is plotted only for the
      $\sigma$-spin species; differences with the $\tau$-spin species
      are negligible in the scales of the figure.}
    \label{fig:Hubbard_Uvar}
  \end{center}
\end{figure}

One can now question whether or not this non-monotonic behavior is stable
when varying the Hamiltonian parameters $t$ and $U$.
In all simulations reported here we fixed energy units by setting $t=1$.
In the limiting case where $U=0$, the fermions in the Hubbard model
are non-interacting, therefore a linear regime in which the current
is always proportional to the driving strength is expected.
In view of these considerations, it is tempting to assume the existence
of a critical value $U^*$ in the Hamiltonian parameters space separating
the linear and NDC behaviors of the current.
As a matter of fact, within numerically accessible system sizes, we observed
NDC only for $U > U^*$, where $U^* \approx 2$.
This can be seen from Fig.~\ref{fig:Hubbard_Uvar}, where we plot
the maximal current drop, measured by
$\langle j \rangle_{f = f^*} - \langle j \rangle_{f = 1}$, as
a function of the on-site interaction strength $U$.
From the inset it is clear that, while for $U \ll 2$ the current
is proportional to the driving, for $U \gg 2$ a bell-shaped behavior emerges.
Of course, on the basis of the data presented in Fig.~\ref{fig:Hubbard_Uvar}
one cannot exclude that $U^\star$ drops with $N$. In this scenario,
in the thermodynamic limit NDC would be observed for any $U>0$;
nonetheless, we point out that a mean-field qualitative argument
given at the end of Sec.~\ref{subsec:magnon} supports
the existence of NDC for $U \gtrsim 2$, thus agreeing with our
findings in Fig.~\ref{fig:Hubbard_Uvar}.
In any case, a significant result of our numerical simulations is
the emergence of  NDC in a physically relevant transport model 
such as the Hubbard model, at small system sizes $N\ge 4$.

\section{Spinless fermion model} \label{sec:tVmodel}

The investigation of the far-from-equilibrium properties of 
the Hubbard model is numerically demanding and an analytical 
treatment appears difficult.
Therefore, in what follows we will focus on a simplified model, 
usually referred to as the $t-V$ model, 
where the spin degree of freedom is neglected.
This model is numerically much more convenient. Moreover,
it will help us in gaining a deeper understanding of the peculiarities of 
charge transport discussed in Sec.~\ref{sec:Hubbard} for the Hubbard model. 

The $t-V$ model considers spinless fermions instead of spin-$1/2$ particles.
Its Hamiltonian reads as follows:
\beq
   \Ham_S = -t \sum_j (c^\dagger_j c_{j+1} + c^\dagger_{j+1} c_j )
   + V \sum_j n_j n_{j+1} \, .
   \label{eq:hamTV}
\eeq
Similarly to 
the Hubbard model~\eqref{eq:Hubbard}, the operators $c^\dagger_j , \, c_j$
create/annihilate a spinless fermion at site $j=1, \ldots , N$ (therefore
they satisfy canonical anticommutation rules), while
$n_j = c^\dagger_j c_j$ is the corresponding number operator.
The system parameters $t$ and $V$ describe, respectively, the nearest neighbor
hopping strength and the fermionic repulsion between contiguous sites.

In direct analogy with what has been discussed for the Hubbard model, we take
open boundary conditions and couple both
ends of the chain to some external baths that inject and extract fermions.
Now the number of Lindblad operators is halved, since we removed
the spin degree of freedom:
\beq
\begin{array}{ll}
   L_1 = \sqrt{\Gamma_L f_L}     \, c^\dagger_1\,, \quad &
   L_2 = \sqrt{\Gamma_L (1-f_L)} \, c_1 \,, \quad \vspace*{1mm}\\
   L_3 = \sqrt{\Gamma_R f_R}     \, c^\dagger_N\,, \quad &
   L_4 = \sqrt{\Gamma_R (1-f_R)} \, c_N \, ;
\end{array}
\label{eq:tVLindblad}
\eeq
the parameters $\Gamma_L$, $\Gamma_R$,
$f_L$, and $f_R$ play roles analogous to those
of the corresponding parameters introduced for spinful fermions 
in Eqs.~\eqref{eq:Lindblad},~\eqref{eq:LindbladRight}.

The Hamiltonian in Eq.~\eqref{eq:hamTV} can be mapped into an
XXZ Heisenberg spin chain plus some spurious contributions, consisting
in external transverse magnetic fields,
which are irrelevant in the isolated (Hamiltonian) case and
do not qualitatively modify the spin current behavior
(see Appendix~\ref{app:mapping}).
Therefore we shall neglect these spurious terms
and concentrate on the XXZ spin-$1/2$ system whose
autonomous Hamiltonian is given by:
\beq
   \Ham_S = \sum_{j=1}^{N-1} \Big[ J_x ( \sigma^x_j \sigma^x_{j+1} +
   \sigma^y_j \sigma^y_{j+1} ) + \ J_z \sigma^z_j \sigma^z_{j+1} \Big] \, ,
   \label{eq:Heisenberg}
\eeq
where $\sigma^\alpha_j$ ($\alpha = x,y,z$) are the Pauli matrices of
the $j$-th spin, and $\Delta \equiv J_z / J_x$ denotes the $xz$ anisotropy;
$N$ is the total number of spins.
In the $t-V$ model language of Eq.~\eqref{eq:hamTV}, the couplings
in Eq.~\eqref{eq:Heisenberg} are given by $J_x = -t/2$ and $J_z = V/4$,
so that the anisotropy $\Delta = -V/2t$.
Strictly speaking, since fermionic interaction between contiguous sites
is repulsive ($V>0$), this would correspond to antiferromagnetic
transverse couplings $J_z >0$. Nonetheless, when considering the
XXZ spin model~\eqref{eq:Heisenberg}, one is not a priori forced by this
constraint and can also analyze the ferromagnetic case $J_z <0$.
Hereafter we set $J_x = 1$ as the system's energy scale.

In the spin-$1/2$ picture, the Lindblad operators~\eqref{eq:tVLindblad}
are mapped into operators flipping the border spins.
Indeed as explained in Appendix~\ref{app:mapping}, apart from
uninfluent phase factors, we have $c^\dagger_j \to \sigma^+_j$ and
$c_j \to \sigma^-_j$.
The Fermi function $f_{L,R}$ is such that $2f_{L,R}-1 \in[-1,1]$ is
the corresponding bath's magnetization per spin in dimensionless units.
As we did for the Hubbard model, we choose 
(with the exception of Sec.~\ref{sec:domainsstability}) a symmetric driving:
$\Gamma_L=\Gamma_R \equiv \Gamma$ and $f_{L,R} = \frac{1}{2} (1 \mp f)$,
so that $f \equiv f_R - f_L \in [0,1]$ 
($f_L\le f_R$, $0\le f_L,f_R \le 1$)
is a single parameter controlling the driving strength.
When $f$ is small we are in the linear response regime, while in the limiting
case $f = 1$ (corresponding to $f_L = 0, f_R = 1$) the left (right)
bath only induces up-down (down-up) spin flips.
The spin current is computed as in Eq.~(\ref{eq:current4}), but without
the contribution of the $\tau$ species:
\begin{equation}
j = J_x (\sigma^x_k \sigma^y_{k+1} - \sigma^y_k \sigma^x_{k+1}).
\label{eq:current5}
\end{equation}

Quantitative numerical and semi-analytical analysis of
the model~\eqref{eq:Heisenberg} are easier than in
the model~\eqref{eq:HeisLadder}. In particular, the local Hilbert space
is halved: for a fixed number of sites $N$,
the size ${\cal N}=2^N$ of a generic state vector describing the system
is decreased by a square root factor with respect to the size 
$(2^N)^2$ of a state vector for 
spin ladder~\eqref{eq:HeisLadder} of length $N$.
It is therefore clear that, without truncating the Hilbert space,
using the Monte-Carlo wave function method one is able to simulate chains
of twice the length of a ladder with the same computational cost. 
We wish to note that in the spin ladder systems we have so far used
only the QT method to perform numerical simulations.
Also the MPO approach could in principle be used, by simply joining
two sites from the opposite spin chains into a single site with a local
dimension $16$ (for a single chain it is $4$).
The complexity of time evolution increases by a factor $4^3$ at a fixed
matrix dimension $D$ (see Appendix \ref{app:numerical} for definitions),
because of singular value decompositions of $16 D\times 16 D$ matrices, 
instead of $4 D \times 4D$.

\section{Spin transport properties in a Heisenberg chain} \label{sec:Heisenberg}

In this section, we study the far-from-equilibrium transport 
properties of the XXZ Heisenberg spin chain~\eqref{eq:Heisenberg},
with the two edge spins coupled to external baths, as described
in the previous section.
We first present, in Secs.~\ref{sec:gapless}-\ref{sec:isotropic},
the results of our numerical simulations, focusing on the NDC
phenomenon and on the appearance of a long-range spin ordering
into ferromagnetic domains. 
Then, in Sec.~\ref{subsec:magnon},
we qualitatively explain our results in terms of one-magnon localization.
Since our findings are suggestive of a phase transition with the
emergence of long-range order, we have also searched for numerical evidence
of such transition by analyzing the spin-spin correlation
function. Our data displayed in Appendix~\ref{app:spinspin}, even if not
conclusive, show a dramatic slowing down of the correlation decay
in the NDC regime, even though
the accessible system sizes are too small for a quantitative analysis 
of a possible phase transition. Finally, in Sec.\ref{subsec:fermionic} we rephrase 
the results in terms of the fermionic current.

Note that the spin current in the XXZ model with antiferromagnetic coupling
is the same as the charge current in the $t-V$ model,
therefore all the results we discuss here for $J_z>0$ also apply
to the case of fermionic transport in Eq.~\eqref{eq:hamTV}.
For the spin transport we also considered cases where $J_z<0$, thus
corresponding to a rather unphysical attractive fermionic interaction
in the $t-V$ model; quite surprisingly, we found that data obtained
for the XXZ chain are insensitive to the sign of the transverse coupling $J_z$.

As far as we know, transport properties of the autonomous model
described by Eq.~\eqref{eq:Heisenberg} have been extensively analyzed 
only within
the linear response regime~\cite{zotos99,prelovsek04,meisner03,narozhny98},
even if a fully comprehensive understanding is still lacking.
In particular, it has been found that the low-temperature and 
the high-temperature 
thermodynamic transport properties are essentially determined
by the $xz$ anisotropy $\Delta$~\cite{zotos99}.
In the zero magnetization sector, the XXZ model is an ideal
conductor for $\vert \Delta \vert <1$, while 
numerical data~\cite{prelovsek04} suggest that the system is 
a normal (diffusive) spin conductor for $\vert \Delta \vert >1$.
The normal conduction in the $\vert \Delta \vert >1$ regime 
has been recently confirmed for systems of much larger size~\cite{prosen},
$N \sim 100$, see also Ref.~\onlinecite{langer08}.
The above two distinct behaviors may be associated 
to two different system phases
at zero temperature: for $-1 \leq \Delta \leq 1$ the system is gapless,
while for $\Delta < -1$ ($\Delta >1$) it is ferromagnetically
(antiferromagnetically) ordered, and the ground state exhibits a finite gap
with the first excited state. 

We now investigate the far-from-equilibrium
properties of the XXZ Heisenberg spin chain, beyond the linear
response regime.

\subsection{Gapless regime} \label{sec:gapless}

\begin{figure}[!h]
  \begin{center}
    \includegraphics[scale=0.34]{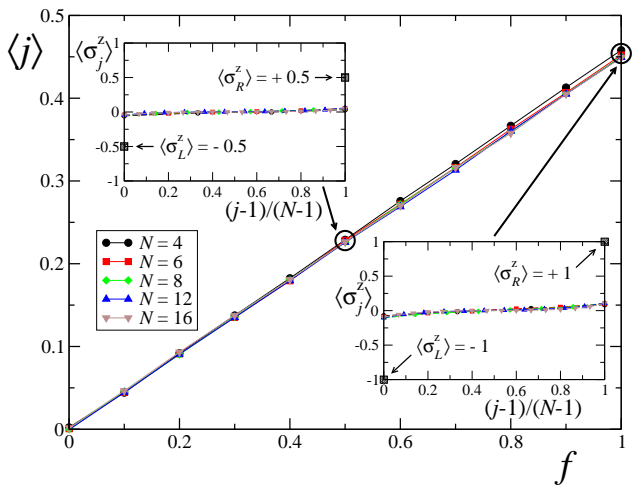}
    \caption{(Color online). Spin current as a function of the driving
      strength for $\Delta = 0.5$.
      The system-bath coupling is set equal to $\Gamma = 1$.
      The insets show spin magnetization profiles versus the scaled spin
      index coordinate, for two values of $f = 0.5, \, 1$.
      Data are obtained from the QT approach.
      }
    \label{fig:Ballistic}
  \end{center}
\end{figure}

We start by considering the gapless phase, where 
the linear response theory predicts ballistic transport.
We found that in the regime $\vert \Delta \vert <1$ the current is always
proportional to the driving and independent of the chain length $N$;
this holds for any value of the driving strength $f \in [0,1]$.
Remarkably, there are no appreciable quantitative differences between
the current evaluated with a ferromagnetic ($J_z < 0$) or
antiferromagnetic ($J_z > 0$) coupling.
Fig.~\ref{fig:Ballistic} displays the spin current
as a function of the driving strength $f$, for $\Delta=0.5$. 
The corresponding spin magnetization profiles for two distinct
values of $f$ are displayed in the insets as a function of the site index.
In both cases we can see a nearly flat profile, that is typical of systems
with ballistic spin propagation~\cite{lepri03};
most importantly, we notice that the stationary spin magnetization
at the borders is very different from the bath magnetizations
$\langle \sigma^z_{L,R} \rangle = \mp f$.

\subsection{Gapped regime} \label{sec:gapped}

The transport properties are much more interesting in the gapped phase
where, in the linear response regime $f\ll 1$, numerical data 
suggest normal diffusive transport~\cite{prelovsek04,prosen}.
For $\vert \Delta \vert >1$ the spin current is no longer monotonic
with $f$ and exhibits a typical bell-shaped behavior, as we observed for the
Hubbard model (see Fig.~\ref{fig:Hubbard_NDC}).

\begin{figure}[!t]
  \begin{center}
    \includegraphics[scale=0.33]{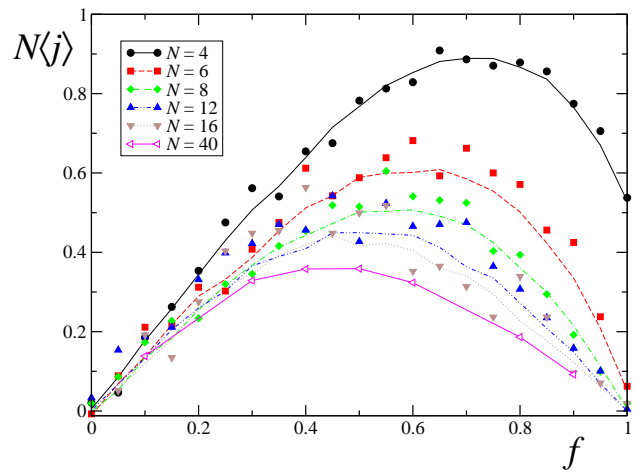}
    \caption{(Color online). Spin current as a function of the driving strength
      $f$, in a chain with anisotropy $\Delta = 2$.
      The system-bath coupling is set equal to $\Gamma = 4$.
      Symbols for $N \leq 16$ are obtained using QT 
      after a simulation time $T = 2.5 \times 10^3$, while
      curves display data for a longer integration time $T= 7.5 \times 10^4$.
      Data for $N=40$ are obtained by MPO, with integration times up to $T=4000$.
      Long-time data for $N \leq 16$ are quoted from Ref.~\onlinecite{paper1}.}
    \label{fig:Insulator_NDC}
  \end{center}
\end{figure}

In Fig.~\ref{fig:Insulator_NDC} we plot the spin current as a function of $f$.
For small $f$ the system behaves as a normal Ohmic conductor, 
as expected from linear response results~\cite{prelovsek04,prosen}: namely
the current increases like $\langle j \rangle \propto f/N$.
After a given value $f^*$ of the driving at which 
the current reaches its maximum,
it starts decreasing with $f$ until it is strongly suppressed at $f = 1$.
Details on the scaling of $\langle j \rangle$ with the system size
at $f=1$ are given in Sec.~\ref{subsec:mu1}.
Here we just point out that, interestingly, NDC is already visible
after short integration times: as a matter of fact, with the 
QT approach (see data for $N \leq 16$),
the characteristic bell-shaped behaviors can be obtained after a
simulation time $T \sim 2.5 \times 10^3$ (symbols) that is much shorter
than the one required to reach the stationary values, $T \sim 10^5$ (curves);
furthermore, NDC features are present even after very short 
times $T\sim 5 \times 10^2$.

The data shown in Fig.~\ref{fig:Insulator_NDC} are 
suggestive of a transition from a normal spin conductor phase at small $f$,
to an insulator phase at large $f$. 
On the other hand, for the achievable system sizes,
the value $f^\star(N)$ where the current reaches a maximum
drops with $N$. Therefore, in principle we cannot exclude an alternative
scenario where at the thermodynamic limit the system becomes an insulator
at any driving strength $f$.
In any case, as far as the bias 
$f$ is increased, a substantial modification
of spin transport properties becomes apparent.
These results are suggestive of a far-from-equilibrium quantum phase transition.
In the light of a recent paper of one of us~\cite{prosen08b},
we have analyzed the spin-spin correlation functions, in order to
see the possible emergence of a phase transition that should be
characterized by the emergence at strong driving strength $f$ 
of a long-range correlation order.
Numerical data, even if not conclusive,
are shown in Appendix~\ref{app:spinspin} and are in support of
such a behavior.

\subsubsection{Behavior at $f=1$} \label{subsec:mu1}

\begin{figure}[!t]
  \begin{center}
    \includegraphics[scale=0.33]{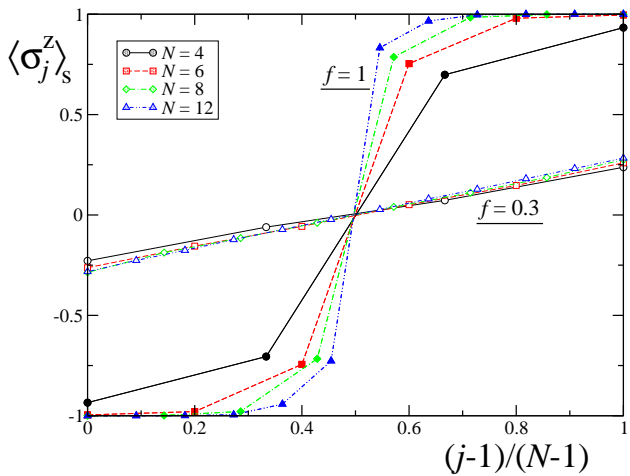}
    \caption{(Color online). Spin magnetization profiles versus scaled spin
      index coordinate for the same parameters as in
      Fig.~\ref{fig:Insulator_NDC}, and driving strengths $f=0.3$
     (empty symbols) and $f=1$ (filled symbols).
     Data are obtained from the QT approach.
      ---Figure acknowledged from Ref.~\onlinecite{paper1}--- }
    \label{fig:SpinZ}
  \end{center}
\end{figure}

As hinted in Ref.~\onlinecite{paper1}, in order to understand the physical
mechanism lying at the basis of NDC, we have to analyze the stationary
spin magnetization profiles. These are shown in Fig.~\ref{fig:SpinZ}.
Note that, in contrast with the fast-time raising up of the NDC phenomenon,
a much longer integration time is required in order to reach a good convergence
for the spin magnetizations, due to the equilibration time scales
that, at $f=1$, 
grow exponentially with the distance of the spin from the chain 
border~\cite{paper1}.

As shown in Fig.~\ref{fig:SpinZ}, magnetization profiles 
in the linear response regime $f \ll 1$ exhibit a constant linear
gradient, where the magnetizations of the two edge spins are close
to the bath magnetizations $\mp f$, as it is expected for normal Ohmic
conductors.
In the limiting case $f = 1$ a peculiar stationary state characterized by
two ferromagnetic domains that are oppositely polarized appears.
Moreover their {\it relative} width increases with the system size.
These domains 
are eventually responsible for strongly suppressing the spin current,
since they inhibit spin flips.
Strictly speaking, evidence of the formation of such domains is also
visible for smaller, though strong drivings, where the NDC effect is
established (see Sec.~\ref{subsec:spinblock}).
We will explain later in Sec.~\ref{subsec:magnon} the physical mechanism
leading to the formation of such domains in the gapped phase.
Here we stress that, as for the gapless regime, we found no quantitative
differences between a ferromagnetic or antiferromagnetic coupling.
This may be somewhat counterintuitive, since in this last case ferromagnetic
domains correspond to a highly excited state for the autonomous system;
the mechanism therefore has its roots
in the genuine far-from-equilibrium dynamics. 

\begin{figure}[!t]
  \begin{center}
    \includegraphics[scale=0.33]{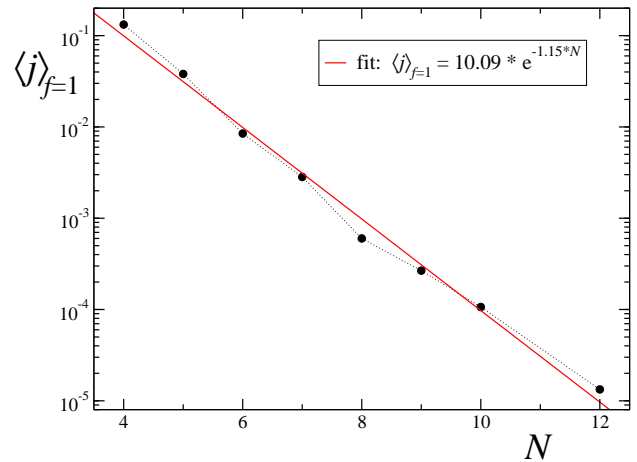}
    \caption{(Color online). Spin current at a fixed driving strength $f=1$
      for the Heisenberg spin model with $\Delta = 2$.
      The system-bath coupling is set equal to $\Gamma=4$; the simulation time
      (QT approach) 
      is $T = 7.5 \times 10^4$. The thick curve displays an exponential fit
      of numerical data (circles).}
    \label{fig:Insulator_Mu1}
  \end{center}
\end{figure}

We already observed that the current for small driving strengths behaves
as for a normal Ohmic conductor:  $\langle j \rangle \propto f/N$.
On the other hand, Fig.~\ref{fig:Insulator_Mu1} shows that at maximum 
bias $f=1$ the current drops to zero exponentially with $N$, that is,
the system is an insulator.
Unfortunately we could not achieve system sizes larger than $N=12$,
since there the current is very small, therefore it would require a huge number
of time steps in order to get reliable results, 
thus making simulations unfeasible
(already for $N=12$ the simulation time $T = 7.5 \times 10^4$ is 
not long enough: the last point in the figure 
has been obtained by observing only one spin flip during the whole 
QT simulation).

\subsubsection{Spin blockade} \label{subsec:spinblock}

At maximum driving the current is strongly suppressed, due to the fact that
ferromagnetic domains of macroscopic length $\sim N/2$ are formed.
Nonetheless, a strong inhibition of the spin current can be also achieved by
only creating a much smaller ferromagnetic region close to each bath.
Indeed, signatures of this spin blockade mechanism are already seen
at $f \sim 0.9 - 0.95$, where asymptotically only a couple of outer spins
reach magnetization values close to $\pm 1$, but still the current is far below
its maximum value $\langle j \rangle_{{f}^\star}$.

\begin{figure}[!t]
  \begin{center}
    \includegraphics[scale=0.33]{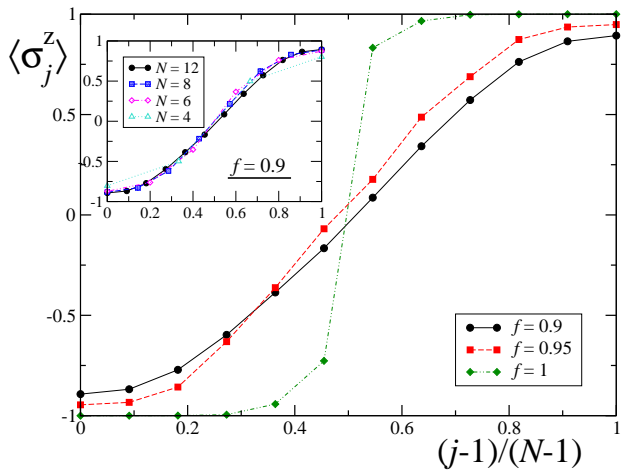}
    \caption{(Color online). Spin magnetization profiles versus scaled
      spin index coordinate
      in the insulating case $\Delta = 2$ with a system-bath coupling
      $\Gamma = 4$, for strong drivings.
      In the main panel we fix a system size $N = 12$ and choose 
      different values of $f$, while in the inset we focus on the case\
      $f = 0.9$ and vary the size according to the legend.
      Data are obtained from the QT approach.}
    \label{fig:SpinBlockade}
  \end{center}
\end{figure}

Some spin magnetization profiles for strong drivings are
explicitly shown in Fig.~\ref{fig:SpinBlockade}.
The dotted-dashed green 
curve corresponds to a maximum driving; there macroscopic
ferromagnetic domains are clearly visible. The other ones are for $f$
slightly less than one, but it is still possible to see that a couple of
spins close to the borders are nearly perfectly down/up polarized.
In the inset we fix $f=0.9$ and vary the system size; we notice that,
when increasing $N$, the number of spins involved in the spin blockade also
increases, in accordance with the results previously shown 
for $f=1$.

\subsubsection{Thickness of the interface region}

\begin{figure}[!t]
  \begin{center}
    \includegraphics[scale=0.33]{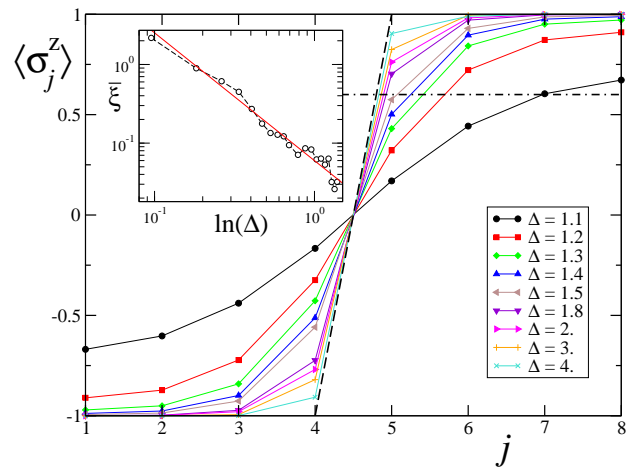}
    \caption{(Color online). Spin magnetization profiles versus scaled
      spin index coordinate for a chain with $N=8$ spins
      at maximum driving ($f = 1$), for different values of $\Delta > 1$;
      the system-bath coupling is set equal to $\Gamma = 1$.
      In the inset we plot an estimate of the thickness of the interface region
      $\overline{\xi}$ as a function of $\ln(\Delta)$; the straight line shows
      the fit $\overline{\xi}\propto (\ln \Delta)^{-1.625}$.
      Data are obtained from the QT approach.}
    \label{fig:Interface}
  \end{center}
\end{figure}

An interesting point that can be addressed is the analysis,
at maximum driving strength $f=1$, of the
characteristic thickness $\xi$ of the interface region,
located around the chain center and dividing the two
ferromagnetic regions.
According to our data shown in Fig.~\ref{fig:Interface}, this size
depends on the system anisotropy $\Delta$. On the other hand, we checked
that dependence on $N$ is negligible.
In order to give a rough estimate of $\xi$, we fixed a threshold
$\overline{\langle \sigma^z_j \rangle}$ (horizontal 
dot-dashed line in the figure, corresponding 
to $\overline{\langle \sigma^z_j \rangle}=0.6$)
and then evaluated the distance $\overline{\xi}$ of the point in each spin
magnetization profile reaching that value from the limiting case 
($\Delta\to\infty$) in which
the chain is exactly split into two ferromagnetic domains
(due to the fact that our problem is on a lattice of finite length, 
the $\Delta\to\infty$ magnetization for $N/2<j<N/2+1$ is estimated after
joining the two perfectly ferromagnetic domains with a skew
dashed line).
In the inset we study the dependence of the distance 
$\overline{\xi}$ on $\Delta$; the straight line shows the
fit $\overline{\xi} \propto (\ln \Delta)^{-1.625}$.
We point out that the semi-analytical argument of the one-magnon
localization length that will be discussed in Sec.~\ref{subsec:magnon}
predicts that the size $\xi$ of the interface region is of the 
order of the one-magnon 
localization length, namely logarithmic in $\Delta$ and $N$-independent.

\subsubsection{Stability of the ferromagnetic domains} \label{sec:domainsstability}

Until now we have only considered situations in which system-reservoir
couplings $\Gamma_{L,R}$ and the driving strengths $\mu_{L,R}$ are symmetric.
At this step one may wonder if the position of the domain wall between the
two ferromagnetic regions in the spin magnetization profiles
is stable against the breaking of such symmetry.
Any imbalance, though small, may in principle cause a shift of the
interface region towards one of the boundaries of the chain.
This would raise some doubts about the stability of the previously
depicted scenario, making our discussion relevant only
for fine-tuned values of the Lindblad parameters.
Below we show that this is not the case.

\begin{figure}[!t]
  \begin{center}
    \includegraphics[scale=0.33]{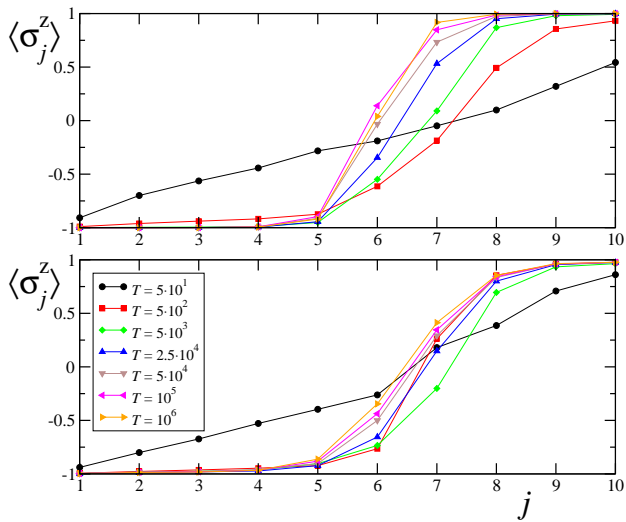}
    \caption{(Color online). Spin magnetization profiles versus
      spin index coordinate for a chain with $N=10$ spins
      and anisotropy $\Delta = 2$, at different integration times $T$.
      Upper panel: symmetric drivings such that $f=1$ and
      system bath couplings $\Gamma_L = 1, \, \Gamma_R = 0.1$.
      Lower panel: the system-bath coupling for the left and the right
      baths is kept fixed and equal to $\Gamma = 1$, while the drivings are
      chosen asymmetrically as $f_L = 0, \, f_R = 0.99$.
      Data are obtained from the QT approach.}
    \label{fig:Domains_asymm}
  \end{center}
\end{figure}

Imbalances of the couplings $\Gamma_{L,R}$ 
have quite tiny effects on the steady state at maximum driving,
as one can see from the upper panel of Fig.~\ref{fig:Domains_asymm}.
We put there a strong asymmetry, by setting
$\Gamma_L = 1, \, \Gamma_R = 0.1$, and thus
admitting a coupling to the right bath
that is one order of magnitude smaller than the one to the left bath.
Differences with respect to the symmetric case are apparent for finite
integration times $T$, where the domain wall is clearly shifted to the right.
On the other hand, as far as $T$ is increased, profiles become more symmetric.
It is not a priori clear whether the steady state is perfectly symmetric,
as those in Fig.~\ref{fig:SpinZ}: from our data we cannot rule out
possible deviations in the position of the domain wall
which are logarithmic in the coupling imbalance; this would
be hardly detectable from a merely numerical analysis.

Stronger modifications are induced by imbalances on the driving
strengths $\mu_{L,R}$. Indeed in that case even a small asymmetry
would cause a weaker spin blockade
on one side, as discussed in Sec.~\ref{subsec:spinblock}
(see the lower panel of Fig.~\ref{fig:Domains_asymm}, in which we put
$f_L = 0, \, f_R = 0.99$);
as a consequence, a deformation and a broadening of the interface region
are also established.
Nonetheless, we point out that these modifications appear to be
continuous in the degree of imbalance,
and thus in principle controllable. 

\subsection{XXX Heisenberg isotropic model} \label{sec:isotropic}

The isotropic XXX Heisenberg chain (that is, $\vert \Delta \vert = 1$) 
corresponds to a limiting, nonetheless interesting situation, since
molecular compounds that are used to investigate one dimensional
spin-$1/2$ transport properties are often very well described by isotropic
antiferromagnetic Heisenberg exchange couplings~\cite{ghirri07,motoyama96}.

\begin{figure}[!t]
  \begin{center}
    \includegraphics[scale=0.34]{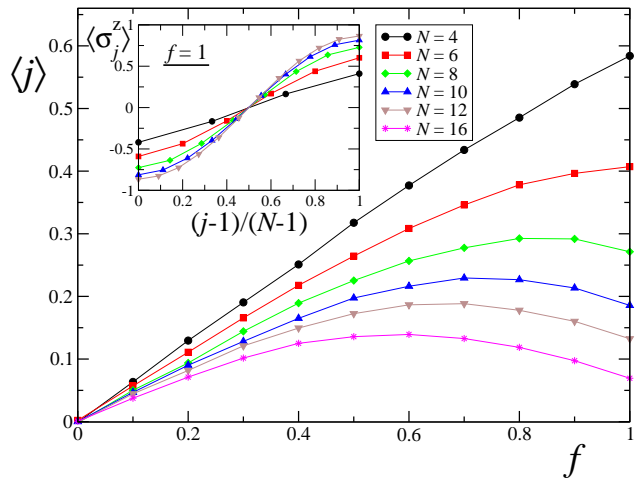}
    \caption{(Color online). Spin current as a function of the driving
      strength for the isotropic case $\Delta = 1$.
      The system-bath coupling is set equal to $\Gamma = 2$.
      The simulation time (QT approach) is $T = 5 \times 10^4$.
      In the inset we show the spin magnetization profiles versus scaled
      spin index coordinate, at maximum driving $f = 1$.
      Different curves stand for various system sizes.}
    \label{fig:Isotropic_NDC}
  \end{center}
\end{figure}

In Fig.~\ref{fig:Isotropic_NDC} we show some numerical data concerning
the behavior of the spin current with respect to the driving field:
NDC is visible only for sufficiently long chains ($N \geq 8$).
A qualitative understanding of this result comes from an analysis of the
spin magnetization profiles at $f=1$, that are plotted in the inset.
As a matter of fact, we can recognize a situation that is similar
to the one already observed at $\vert \Delta \vert > 1$ and $f \lesssim 1$
(see Fig.~\ref{fig:SpinBlockade}): by increasing $N$, a partial spin
blockade of the outermost spins is established. This progressively inhibits
the current flow along the chain, thus setting up the NDC mechanism.
Notice also that, at small $N$ values ($N = 4, \, 6$), the spin blockade
is very weak; this prevents systems of very small size from exhibiting the 
NDC phenomenon, even though a nonlinear dependence of the spin
current on the driving strength can already be seen.

\begin{figure}[!t]
  \begin{center}
    \includegraphics[scale=0.31]{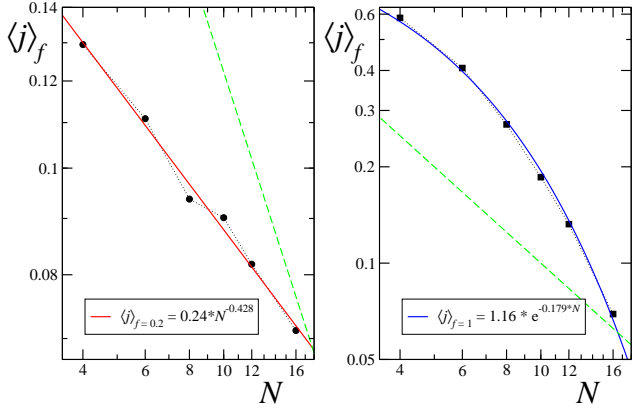}
    \caption{(Color online).
      Spin current for $\Delta=1$ at a fixed driving strength,
      $f=0.2$ on the left,
      while $f=1$ on the right (see Fig.~\ref{fig:Isotropic_NDC}).
      Full curves are fits of numerical data (obtained from QT): while
      an inverse power-law fit works well at $f \ll 1$, for maximum driving
      an exponential fit seems adequate. The dashed lines indicate a
      behavior $\langle j \rangle_f \sim 1/N$ and are plotted as guidelines.}
    \label{fig:Isotropic_Mu1}
  \end{center}
\end{figure}

We conjecture that also for the isotropic XXX chain
the NDC phenomenon is stable at the thermodynamic limit.
Indeed, our simulations suggest the following picture:
on one hand, at small $f$ the current decreases as 
$\langle j \rangle \propto N^{-\alpha}$ with $\alpha \approx 0.4$
(data for $f=0.2$ are shown in Fig.~\ref{fig:Isotropic_Mu1}, left;
we checked that this is consistent for all $f \lesssim 0.3$).
On the other hand, at $f = 1$ it drops to zero faster than linearly with $N$
(probably exponentially, as suggested by Fig.~\ref{fig:Isotropic_Mu1}, right),
thus indicating a relative current drop 
$1 - \langle j \rangle_{f=1} / \langle j
\rangle_{f=f^*}$ that increases with $N$ towards the unit value;
here $f^*$ denotes the driving at which the current is maximal.
In both panels a log-log scale has been used, so to make visible
the distinction between a power-law scaling at small $f$ and
an exponential behavior at $f = 1$.
We also plotted the $1/N$ behavior (dashed lines) expected
for normal Ohmic conductors, such to show that
for $f \ll 1$ the decay is slower than that, and for $f=1$ it is faster).

\subsection{One-magnon localization} \label{subsec:magnon}

The formation of ferromagnetic domains and the negative differential 
conductivity phenomenon in the gapped regime $\vert \Delta \vert >1$
can be qualitatively explained in terms of localization of one-magnon
excitations created at the borders of a ferromagnetic domain.
We should immediately point out that the following argument is independent
of the ferromagnetic of antiferromagnetic coupling.
This reflects into the fact that the steady state with ferromagnetic
domains is completely driven by dynamical effects, and not by the
system ground state, being actually an antiferromagnet for $J_z>1$.

Given a ferromagnetic state
$\ket{0} \equiv \ket{\downarrow \downarrow \cdots \downarrow}$,
one-magnon excitations have the general form
$\sum_{k=1}^N \alpha_k \ket{k}$, where $\ket{k}=\sigma^+_k \ket{0}$
describes the state with the $k$-th spin flipped.
If the autonomous 
XXZ chain has open boundary conditions, there is an energy gap $2|J_z|$
between the states $\ket{1}$ and $\ket{N}$
(spin-flip excitations at the boundaries) and the states
$\ket{2}$, $\ket{3}$,..., $\ket{N-1}$.
Indeed, we have 
\beq
   \begin{array}{l}
   \displaystyle \langle 0 | \Ham_s | 0 \rangle = (N-1) J_z, \\
   \displaystyle \langle 1 | \Ham_s | 1 \rangle =
                 \langle N | \Ham_s | N \rangle = (N-3) J_z, \\
		 \langle 2 | \Ham_s | 2 \rangle = ...
	       = \langle N-1 | \Ham_s | N-1 \rangle = (N-5) J_z,
   \end{array}
   \label{eq:surface1}
\eeq
where $\Ham_s$ is the XXZ Hamiltonian~\eqref{eq:Heisenberg}.
Only nearest-neighbor spin-flipped states are coupled 
and the coupling strength is $2|J_x|$:
\beq
   \langle k | \Ham_s | k+1 \rangle = 2 J_x,\quad k=1,...,N-1.
   \label{eq:surface2}
\eeq
As shown in Appendix~\ref{app:magloc}, the autonomous 
model~\eqref{eq:surface1}-\eqref{eq:surface2} is exactly solvable in the 
limit of large $N$ and
spin-flip excitations created at the borders of the chain remain 
exponentially localized when $|J_z|/|J_x|=|\Delta|>1$, over a 
localization length $\ell\sim 1/ \ln \vert \Delta \vert$.

\begin{figure}
  \begin{center}
    \includegraphics[scale=0.32]{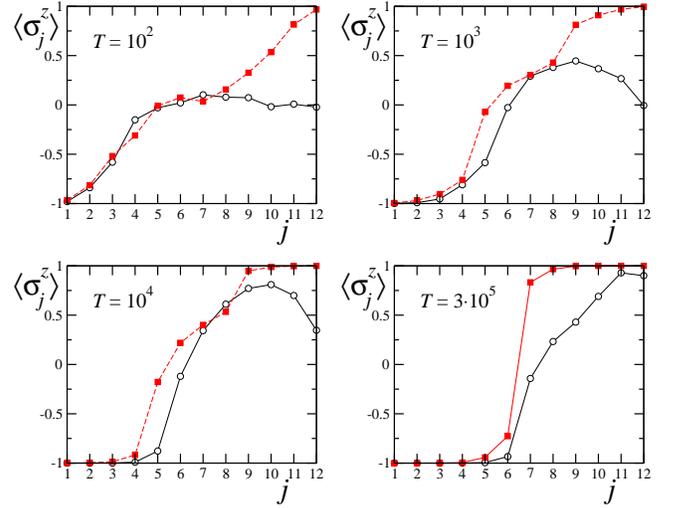}
    \caption{(Color online). Spin magnetization profiles at
      $\Delta = 2$, $\Gamma = 4$
      for a maximal driving strength: $f = 1$.
      The various panels correspond to different times $T$ 
      (in QT simulation) at which
      the profiles are plotted.
      Circles are for a single bath coupled to the first spin ($j=1$),
      while squares are for two baths at the ends of the chain.}
    \label{fig:SpinZ_profiles}
  \end{center}
\end{figure}

We now consider the coupling to external baths. 
First, it is instructive to discuss the case in which the system
is coupled to a single, fully polarized reservoir, $f_L=0$.
Regardless of the anisotropy $\Delta$, the stationary state
is pure and ferromagnetic, namely
$\ket{\downarrow \downarrow \cdots \downarrow}
\bra{\downarrow \downarrow \cdots \downarrow}$,
since the Hamiltonian~\eqref{eq:Heisenberg} conserves the overall magnetization
while at the left boundary of the chain only the lowering operator
$L_2\propto \sigma_1^-$ acts (note that the convergence to
the stationary ferromagnetic state can be rigorously proven 
following Ref.~\onlinecite{vittorio}).
As shown in Fig.~\ref{fig:SpinZ_profiles} (see circles),
the time scale required for the convergence of the $j$-th spin to the
equilibrium state $\ket{\downarrow}$ scales exponentially with $j$. 
Consider now an intermediate state with $m$ spins down.
To enlarge the ferromagnetic domain, one-magnon excitations should be
propagated, through $\sigma^x_j \sigma^x_{j+1}$ and $\sigma^y_j \sigma^y_{j+1}$
exchange couplings of the Hamiltonian~\eqref{eq:Heisenberg},
across the ferromagnetic domain to the left chain boundary.
Suppose, for instance, that we have the leftmost $m$ spins down and
the $(m+1)$-th spin up, and that this excitation propagates to the left bath;
then the bath can flip this spin down, thus ending up with a ferromagnetic
domain with $m+1$ leftmost spins down.
The crucial point is that the one-magnon propagation is exponentially
localized at $|\Delta|>1$. Hence, exponentially long time scales
$\propto \exp(j / \ell) \sim \exp(j \ln \vert \Delta \vert)$ are required
to polarize the $j$-th spin.

Consider now two baths at $f=1$. Due to exponential one-magnon 
localization, essentially only the nearest bath is felt by the spins 
in the chain, with the exception of an interface region between the two
ferromagnetic domains, whose length is of the order of the one-magnon 
localization length. As shown in Fig.~\ref{fig:SpinZ_profiles}
(squares), apart from the interface region, the spin magnetization profiles
for the spins closer to the left than to the right bath essentially evolve
as in the single-bath case.

These ferromagnetic domains are responsible for strongly inhibiting
spin flips, and therefore for suppressing the spin current at $f=1$.
Since at small $f$ the current grows linearly, we can conclude that,
due to the continuity of $\langle j \rangle_{f}$, a region of 
negative differential conductivity exists.
Finally, we note that, in agreement with our numerical data, the one-magnon
argument does not distinguish between ferromagnetic ($J_z<0$)
and an antiferromagnetic ($J_z>0$) spin couplings.

The argument developed in this section can be extended 
to the spin ladder model~\eqref{eq:HeisLadder} obtained 
by applying a Jordan Wigner Transformation (JWT) to the Hubbard Hamiltonian.
Assume that we have a ferromagnetic domain of the $m$ spins of both
species ($\sigma$ and $\tau$) closest to, say, the left bath.
Now consider a spin flip for the $(m+1)$-th spin, for instance
of the $\sigma$ species
and assume that the spins of the $\tau$ species can be treated
within the mean-field approximation.
The only energy term not constant for the autonomous
evolution~\eqref{eq:HeisLadder}
restricted to the one-magnon sector ($\sigma$ species) 
is $\frac{U}{4}\sum_j \sigma_j^z \tau_j^z$.
Within mean-field approximation, we substitute $\tau_j^z \to 
\langle \tau_j^z \rangle$,
and $\langle \tau_j^z \rangle =1$ for $i=1,...,m$, 
while $\langle \tau_{m+1}^z \rangle=0$.
There is an energy gap $\frac{|U|}{2}$ between states with the spin flipped
belonging to sites from $1$ to $m$ and the state with the 
$(m+1)$-th spin flipped.
The hopping strength is $|t|$. Therefore magnon localization as in
the XXZ model takes place for $|U|>2|t|$.
Such prediction is compatible with the numerical results shown in 
Fig.~\ref{fig:Hubbard_Uvar},
where NDC for the Hubbard model is observed only for $U> U^\star$,
with $U^\star\approx 2$. However, the argument is of a mean field
nature and therefore has to be considered weaker than the one developed for 
the XXZ chain.

\subsection{Discussion in terms of the charge current in the fermionic model}
\label{subsec:fermionic}

At this stage, it is useful to summarize the main results obtained for 
the spin chain in terms of the original electronic $t-V$ model.
Since $\frac{1}{2}(1+ \langle \sigma_k^z \rangle)$ corresponds 
to the electronic charge density at site $k$, the ferromagnetic 
domains observed at $f=1$ corresponds, in the fermionic picture,
to a phase separation, with all the electrons frozen in the right
half of the lattice, close to the emitter electrode. Therefore, charge
transport is inhibited, provided that $V/t>2$. Note that this charge 
clustering takes place in spite of the repulsive nature of
electron-electron interactions. The magnon localization argument can be 
straightforwardly reformulated in terms of the fermionic model.
In particular, a one-magnon excitation becomes a single electron (hole)
propagating on an empty (filled) lattice. The single-bath case 
can then be interpreted in terms of depletion (filling) of the 
lattice by means of a single lead, playing the role of a 
charge collector (emitter). This process requires exponentially
long time scales at $V/t >2$. 
Finally, we point out that the NDC regime is observed for strongly 
interacting systems ($V/t>2$) and in the far-from-equilibrium regime,
corresponding to large bias voltages 
$eV\gg k_BT$, so that the Fermi functions for the collector and
the emitter electrodes, evaluated at the energy differences
$E_1$ and $E_N$, satisfy $f_L\approx 0$ and $f_R \approx 1$,
respectively.

\section{Nonintegrable model: staggered magnetic field} \label{sec:stagger}

The high-temperature transport properties in one-dimensional
quantum many-body systems are strongly affected by the presence
of conservation laws~\cite{castella95,zotos96,zotos97,saito96}.
In particular, the existence of local conserved quantities  
$Q_n$, $n=1,2,\ldots$, typically leads to an ideally
conducting - ballistic behavior, at all the temperatures.
This is a consequence of the inequality due to Mazur \cite{mazur} which bounds
the time averaged current-current autocorrelation function as
$\lim_{t\to\infty}(1/t)\int_0^t {\rm d}t' \ave{J(t')J(0)}_\beta \ge
\sum_n |\ave{J Q_n}_\beta|^2$ where
$\ave{X}_\beta\equiv\tr[\exp(-\beta H) X]/\tr[\exp(-\beta H)]$ and $Q_n$ are chosen
and normalized such that $\ave{Q_n Q_m} = \delta_{n m}$.
As one can see, this argument essentially depends on, first, the existence
of nontrivial conserved quantities (as is typically the case {\em only}
for completely integrable systems), and, second, on the overlaps $\ave{Q_n J}_\beta$
between the conservation laws $Q_n$ and the transporting current $J$ in question.
For example, for the XXZ model all the conservation laws have zero overlaps 
with the spin current, 
$\ave{J_s Q_n} = 0$, which allows in the gapped regime $\vert \Delta \vert> 1 $
for the normal diffusive spin transport as discussed earlier, whereas, for example,
the heat transport is ballistic as the energy current is just one of the 
conserved quantities $J_E = Q_3$ \cite{zotos97}.
However, even though the XXZ Heisenberg model is completely integrable,
we have seen that the spin transport properties in the  
far-from-equilibrium regime are very different from the linear
response regime behavior. 
Therefore, we might expect that the presence of NDC is not 
related to integrability of the Heisenberg model.

\begin{figure}[!t]
  \begin{center}
    \includegraphics[scale=0.34]{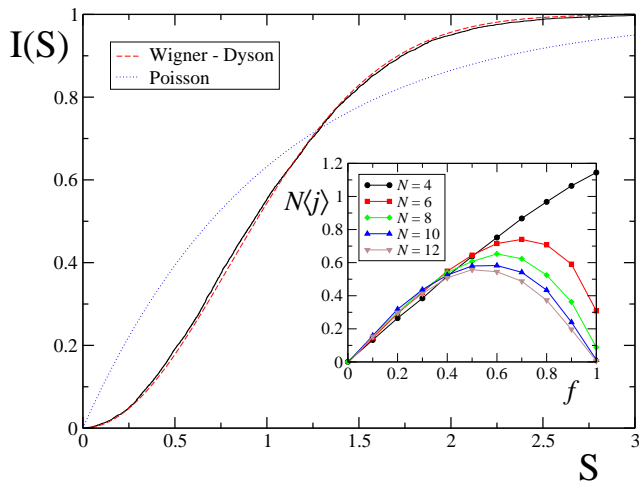}
    \caption{(Color online). Integrated level spacing distribution $I(S)$ for
      the Hamiltonian in Eq.~\eqref{eq:HeisStag} with $\Delta=1.3$, $B = 0.3$,
      $N=16$; data (full black curve) correspond to the zero magnetization
      sector.
      In order to apply the random matrix theory, the system Hamiltonian
      has to be diagonalized in a subspace in which no symmetries
      (except the time-reversal) are left.
      For this purpose, in addition to fixing the total magnetization,
      we applied the staggered magnetic field
      on all spins but the first one (i.e., we supposed that the last sum
      in Eq.~\eqref{eq:HeisStag} runs from $2$ to $N$).
      This breaks the spatial reflection symmetry $j \to N-j$,
      without affecting the transport properties under investigation.
      Red dashed (blue dotted) curve indicates Wigner-Dyson (Poissonian) 
      statistics, typical for chaotic (integrable) systems.
      In the inset (data from QT) 
      we plot the spin current as a function of the driving
      strength, for different chain lengths; we fixed a system-bath
      coupling $\Gamma = 1$.}
    \label{fig:LSS}
  \end{center}
\end{figure}

In order to check the stability of the nonlinear transport highlighted in
the previous sections with respect to breaking system integrability,
we considered a slightly modified spin chain model, in which
a staggered magnetic field along the $z$ direction is added to the
Heisenberg Hamiltonian of Eq.~\eqref{eq:Heisenberg}.
Namely, we studied the following autonomous model:
\beq
   \Ham_S  =  \displaystyle \sum_{j=1}^{N-1} \Big[  (\sigma^x_j \sigma^x_{j+1}
   + \sigma^y_j \sigma^y_{j+1} ) + \Delta \sigma^z_j \sigma^z_{j+1} \Big]
   - B \sum_{j=1}^N (-1)^j \sigma^z_j \, .
   \label{eq:HeisStag}
\eeq

\begin{figure}
  \begin{center}
    \includegraphics[scale=0.33]{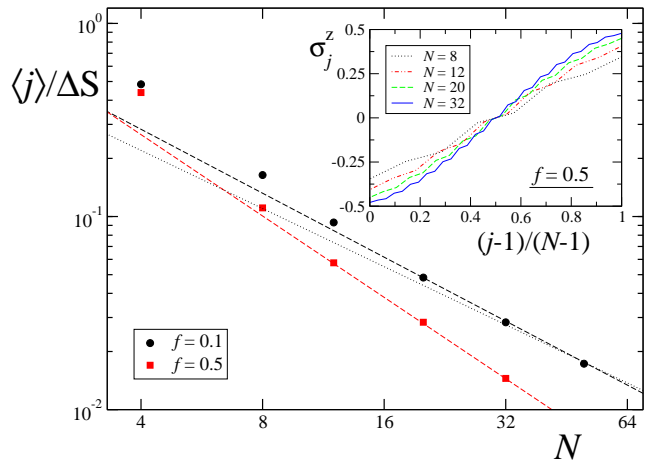}
    \caption{(Color online). Dependence of the scaled spin current on the
      chain length in the staggered Heisenberg model, with $\Delta = 1.3$, $B=0.3$, 
      and $\Gamma = 1$
      ($\Delta S$ is the magnetization difference across the chain),
      at driving strength $f=0.1$ (circles) and $f=0.5$ (squares).
      The two dashed lines respectively denote the behaviors $\sim 1.3/N^{1.1}$,
      (for circles) and $\sim 1.85/N^{1.4}$ (for squares), whereas dotted line
      indicates $1/N$ behavior found in the linear response regime. 
      In the inset we show magnetization profiles for $f=0.5$ and 
      different system sizes.
      Data are obtained from MPO.}
    \label{fig:Curr_Stag_NDC}
  \end{center}
\end{figure}

Interestingly, the model~\eqref{eq:HeisStag} exhibits a transition
from integrability to quantum chaos when increasing the field strength $B$.
This can be detected in the change of the spectral statistics of the 
system~\cite{haake:BOOK,guhr}.
In particular, in Fig.~\ref{fig:LSS}
we plot the integrated level spacing distribution $I(S)$ 
[$I(S)$ is the probability that a randomly chosen level spacing 
-- normalized to the mean level spacing -- is less than $S$].
It is shown that, for a given set of
parameters and $B \neq 0$, the level statistics follows
the universal predictions of the random matrix theory (Wigner-Dyson 
statistics in presence of time-reversal symmetry),
as typical for chaotic (strongly non-integrable) systems.
In the inset we analyzed the corresponding spin current as a function
of the driving strength: we found a qualitatively analogous behavior
as in the integrable case, with a NDC regime still clearly visible.
We also checked the presence of a normal Ohmic conduction for small drivings:
from Fig.~\ref{fig:Curr_Stag_NDC} one can see that,
for $f=0.1$, the spin current scales as $\langle j \rangle \propto 1/N$,
according to Ohm's law of diffusive transport (see also Ref.~\onlinecite{prosen}),
therefore spin transport is normally diffusive, as expected for a chaotic system.
In analogy with the integrable case, for larger gradients where the current
behaves highly nonlinearly, the spin current decays faster
than linearly 
(at $f=0.5$ we found a decay $\langle j \rangle \sim \Delta S/N^{1.4}$,
thus indicating an insulating behavior in the thermodynamic limit).
Also magnetization profiles in that case are not linear anymore,
and the emergence of a weak spin blockade is already visible at $f=0.5$,
for sufficiently large sizes (see the inset of Fig.~\ref{fig:Curr_Stag_NDC}),
similarly to what has been already discussed in Sec.~\ref{subsec:spinblock}.

\begin{figure}
  \begin{center}
    \includegraphics[scale=0.33]{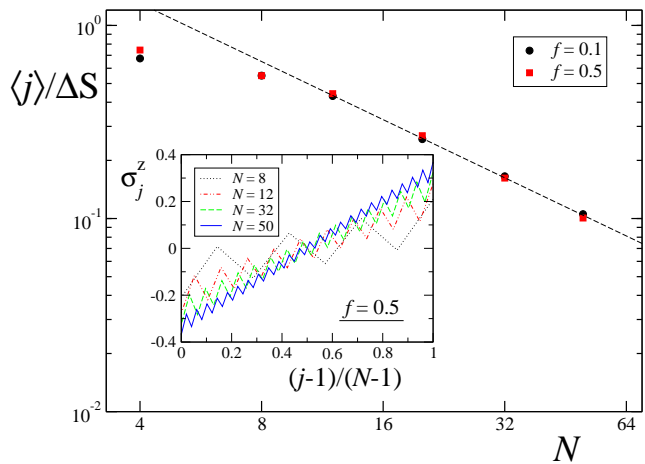}
    \caption{(Color online). Dependence of the scaled spin current on the chain
      length for the staggered Heisenberg model, with $\Delta  = 0.5$, $B=0.5$,
      and $\Gamma = 1$.
      Both for strong and weak drivings the system shows 
      a normal Ohmic behavior, as
      it is depicted by the straight dashed line, which corresponds
      to $ \sim 5.2/N$. The magnetization has an approximately linear profile,
      as shown in the inset, with a slight even-odd oscillation that is 
      due to a staggered magnetic field.
      Data are obtained from MPO.}
    \label{fig:Curr_Stag_Ohm}
  \end{center}
\end{figure}

On the other hand, we considered a situation in which the system does not
exhibit NDC ($\Delta = 0.5$, $B=0.5$), but still it is quantum chaotic with 
respect to energy level statistics.
In that case we found the usually predicted behavior for non integrable
systems~\cite{castella95}:
the current is always proportional to the driving for any value of $f$,
and a normal Ohmic regime for both small and strong drivings emerges,
as shown in Fig.~\ref{fig:Curr_Stag_Ohm}.
This comes in sharp contrast to the integrable case, where 
at $\Delta<1$ ballistic spin transport takes place.
Consistently with normal metallic conductors, the spin magnetization profiles
exhibit a linear gradient (see the inset of Fig.~\ref{fig:Curr_Stag_Ohm}).

From the results of this section we can conclude that the NDC phenomenon
is governed by the anisotropy parameter $\Delta$ and is insensitive to 
the transition from integrability to quantum chaos. 
The appearance of NDC in the quantum chaos regime is particularly significant 
since in this case normal Ohmic conduction is expected in the linear response
regime (small driving $f$). Due to the insulating behavior 
observed for $\vert \Delta \vert > 1$ at
large driving strengths $f$, the system should undergo a metal-to-insulator
transition when increasing $f$.

\section{Conclusions} \label{sec:conclusions}

In summary, we have performed extensive numerical simulations showing that the
one dimensional Hubbard model, as well as the corresponding spinless fermion
model, exhibits a regime of negative differential conductivity,
in which the particle current decreases as one increases the driving field.
The same phenomenon, translated in terms of the spin current,
is observed and studied also in the anisotropic Heisenberg
spin chain.
Our numerical data show that in the Ising-like regime,
corresponding to anisotropy parameter $\vert \Delta \vert > 1$,
the system is a normal conductor for small driving fields, while it displays
negative differential conductivity for stronger drivings.
The phenomenon is stable against the breaking of integrability, as, for instance,
it persists in the presence of a staggered magnetic field.
On the other hand, in the XY-like regime, $\vert \Delta \vert <1$, ballistic
conduction with current proportional to driving field without negative
differential conductivity is observed at all drivings.
Negative differential conductivity is schematically explained using
the spectrum of one-magnon excitations,  which become localized for
$\vert \Delta \vert > 1$ at the chain boundaries.

The observed negative differential conductivity 
arises as an outcome of a beautiful interplay between
coherent many-body quantum dynamics of the 
interacting electrons/spin chain and incoherent
charge injection-extraction/spin pumping operated by 
macroscopic leads/spin baths.
While our results are suggestive of a metal-insulator
phase transition when driving our model systems far from
equilibrium, new analytical approaches are required 
to ascertain whether NDC survives at the thermodynamic limit. 
At any rate, the discussed magnon-localization mechanism 
for NDC is of potential interest for nanoscale devices such as 
current/heat diodes and transistor.

{\it Note added.} After the completion of this work we became aware
of a related paper~\cite{tasaki09} where it is shown
that negative differential conductivity can be also achieved in
a continuum limit of a fermionic tight binding model,
driven far from equilibrium by embedding it between
two free fermionic reservoirs.

\section*{Acknowledgments}

We thank G. Falci for useful discussions.
TP and M\v Z acknowledge support by the program P1-0044 of Slovenian Research
Agency.

\appendix

\section{Numerical methods} \label{app:numerical}

\subsection{Quantum trajectories approach} \label{sec:qtraj}

As compared to full density matrix simulations, the advantage of a quantum
trajectory approach is that, instead of storing and evolving a density matrix
of size ${\cal N} \times {\cal N}$, one works with a stochastically evolving
state vector of size ${\cal N}$ ($\mathcal{N}$ being the Hilbert space
dimension of the system).
The first two terms in the r.h.s. of Eq.~\eqref{eq:MasterEq} can be regarded as
the evolution performed by an effective non-Hermitian Hamiltonian
$\Ham_{\rm eff} \equiv \Ham_S + i {\cal K}$, with
${\cal K} = - \frac{1}{2} \sum_m L^\dagger_m L_m$; the last term is responsible
for the so-called quantum jumps, as explained below.
If the initial density matrix describes a pure state
$\rho(t_0) = \ket{\phi(t_0)} \bra{\phi(t_0)}$, after a small amount of time
$dt$ it evolves into the statistical mixture
\beq
   \rho(t_0 + dt) = \Big( 1 - \sum_m dp_m \Big) \ket{\phi_0} \bra{\phi_0}
   + \sum_m dp_m \ket{\phi_m} \bra{\phi_m},
\eeq
where $dp_m=\langle \phi(t_0)\vert L_m^\dagger L_m \vert \phi(t_0) \rangle dt$
and the new states are defined by
\beq
   \ket{\phi_0} = \frac{e^{- i \Ham_{\rm eff} dt} \ket{\phi(t_0)}}
   {\sqrt{1 - \sum_m dp_m}} \, , \quad
   \ket{\phi_m}=\frac{L_m \ket{\phi(t_0)}}{\lVert L_m \ket{\phi(t_0)}\rVert}\,.
\eeq
Therefore, with probability $dp_m$ a jump to the state $\ket{\phi_m}$ occurs,
while with probability $1-\sum_m dp_m$ there are no jumps and the system
evolves according to $\Ham_{\rm eff}$.

In practice, one starts from a pure random state $\ket{\phi(t_0)}$ and, at
intervals $dt$ much smaller than the relevant dynamical time scales,
a random number $\eps \in [0,1]$ is chosen.
If $\eps \leq \sum_m dp_m$, the state of the system jumps
to one of the states $\ket{\phi_m}$ (to $\ket{\phi_1}$ if $0 \leq \eps
\leq dp_1$, to $\ket{\phi_2}$ if $dp_1 < \eps \leq dp_1+dp_2$, and so on).
On the other hand, if $\eps > \sum_m dp_m$ the evolution with the
non-Hermitian Hamiltonian $\Ham_{\rm eff}$ takes place, thus ending up
in the state $\ket{\phi_0}$.
This process has to be repeated as many times as $n_{\rm steps} = T / dt$,
where $T$ is the total evolution time.
Assuming that there exists a single out-of-equilibrium steady state
$\rho_s$, the expectation values $\langle A \rangle=\tr(A\rho_s)$
of any observable $A$ are obtained after averaging in time
$\langle \phi(t) | A |\phi(t)\rangle$ up to a long enough time $T$, skipping
the initial {\em convergence time} $T_{\rm conv}$ needed for a solution
of the Lindblad equation to converge into the stationary one.
 
In the cases discussed in this paper, the non-Hermitian evolution of
$e^{-i \Ham_{\rm eff} dt}$ has been simulated with a second order
time Trotter expansion; a time slicing
$dt \sim 2 \times 10^{-3}$ is sufficient to obtain accurate results.
The fermionic systems taken into account are mapped into lattice spin models;
simulations are then directly performed on spin-$1/2$ Hamiltonians
[see Eqs.~\eqref{eq:HeisLadder} and~\eqref{eq:Heisenberg}].
We were able to simulate systems with sizes up to ${\cal N} \approx 10^5$
(that is, $N=\log_2{\cal N} \approx 16$ spin-$1/2$ particles).
We calculated the stationary spin current $\langle j \rangle$
on the basis on Eq.~(\ref{eq:current2}). 
That is, we computed $\langle j \rangle$
by summing up all down-up flips minus all up-down flips at the right end 
of the chain or all up-down flips minus all
down-up flips at the left end, and then dividing by the simulation time.
In the fermionic language, this corresponds to 
counting the number of electrons
that enter the system minus the number of electrons that leave
the system per unit time through the right electrode (emitter) or 
the number of electrons
that leave the system minus the number of electrons that enter
the system per unit time through the left electrode (collector).
We also checked that the current obtained in this way is,
up to statistical fluctuations due to finite integration times, 
equal to the one computed through 
Eqs.~(\ref{eq:current4}), (\ref{eq:current5}).
A good convergence for $\langle j \rangle$
is already reached at $T_{\rm conv} \sim 10^4$, while integration
times $T$ of one order of magnitude longer are required in order to determine
the stationary magnetization profiles (e.g., $T\sim 3 \times 10^5$
for $N=12$ in Fig.~\ref{fig:SpinZ}).

Data presented for the Hubbard model and the spinless fermion model
with $N \leq 16$ have been obtained with this approach.
For further details of the implementation of 
the quantum trajectories approach see Ref.~\onlinecite{carlo04}.

\subsection{Matrix Product Operator formalism} \label{sec:mpo}

We have used the MPO ansatz for spin-1/2 particles. 
An arbitrary density matrix $\rho$ of a chain of $N$ spins 1/2 can be 
expanded over all possible products of Pauli operators forming a basis of a 
$4^N$ dimensional Hilbert space of operators:
\begin{equation}
  \ket{\rho} = \sum_{\vec{s}}  \alpha_{\vec{s}} \ket{\sigma^{\vec{s}}} \, ,
  \label{eq:superket}
\end{equation}
where we used the compact notation
$\sigma^{\vec{s}}=\sigma_1^{s_1} \cdots \sigma_{N}^{s_{N}}$,
$\vec{s}\equiv s_1 \cdots s_N$, and $s_i \in \{0,1,2,3\}$, with
$\sigma^0=\mathbbm{1}, \sigma^1=\sigma^x, \sigma^2=\sigma^y,\sigma^3=\sigma^z$,
and where lower indices in the Pauli operators denote the site number of 
the spin on which it operates. 
The use of ket notation
in $\ket{\rho}$ outlines the fact that the density matrix
$\rho$ can be seen as a vector in the operator Hilbert space spanned 
by the basis vectors $\ket{\sigma^{\vec{s}}}$.
In the MPO ansatz, the expansion coefficients $\alpha_{\vec{s}}$ are 
expressed as traces of products of $N$ $D \times D$ dimensional matrices
$\mm{A}_i^{s_i}$, $i=1,\ldots,N$, as
\begin{equation}
  \alpha_{\vec{s}} = \tr{(\mm{A}_1^{s_1} \cdots \mm{A}_{N}^{s_{N}})} \, .
  \label{eq:MPO}
\end{equation}
Thus, a (density) operator (\ref{eq:superket}) is completely specified and thus
parametrized in terms of a set of $4N$ matrices $\mm{A}_i^{s_i}$,
four for each site $i$, $s_i=0,1,2,3$. We note that these matrices are not
related to physically observable quantities. 
The propagator corresponding to the master equation~\eqref{eq:MasterEq}
is written as a product of propagators for small steps of length $dt$,
typically $dt = 10^{-1}$.
Each small time-step propagator is then split using a third order Trotter 
expansion into parts composed of mutually commuting two-spin terms. 
These nearest neighbor two-spin terms are then basic transformations 
performed within the MPO ansatz, namely after each such two-spin transformation,
a singular value decomposition is performed in order to restore the shape of the
ansatz~\eqref{eq:MPO}. However, this step has to be combined with a truncation
of the resulting matrices to a smaller, fixed dimension $D$.
Dimension $D$ is then chosen as a parameter by which we control the accuracy
of the method.
Note that the minimal necessary dimension $D$ is related to the bipartite 
entanglement of $\ket{\rho}$ in the Hilbert space of operators. This implies
that MPO method will fail if the state $\rho$ builds up strong quantum
correlations over large distances. For a review on MPO techniques
see Ref.~\onlinecite{murg}, while details of the implementation
of the MPO method in quantum master equations can be found in
Ref.~\onlinecite{prosen}.

Starting from an arbitrary initial
density matrix $\rho(t_0)$ we are interested in the asymptotic nonequilibrium
steady state $\rho_s$ reached after a sufficiently long time
of simulation, i.e., of relaxation.
Once $\rho_s$ is obtained, various expectation values can 
be evaluated. 
In particular, the spin current is obtained from Eq.~(\ref{eq:current5}) as
$\langle j \rangle= 
J_x \tr[ \rho_s (\sigma^x_k \sigma^y_{k+1} - \sigma^y_k \sigma^x_{k+1})]$.
The simulation time that is required to reach a stationary
$\rho_s$ in the regime of NDC may greatly depend on the 
imposed gradient (driving field). 
In the most difficult situations that we studied here, the relaxation time
after which the local current inhomogeneities decreased to a few percent
was $T \sim 4000$. 
Note that this relaxation time is not directly comparable to the time 
needed in the quantum trajectories approach, as there large $T$ 
is needed also to perform statistical averaging. In the MPO approach 
averaging over Hilbert space is exact, up to the truncation imposed by 
finite matrix dimension $D$.
The main advantage of MPO as compared to other approaches is that it generally
enables simulation of larger systems.
For small drivings one can go up to $N \sim 100$,
while in the present work, where systems are driven far from equilibrium, 
we could reach sizes $N \sim 40$.

Data shown for the spinless fermion model with $N > 16$
have been obtained with MPO technique.

\section{Mapping fermionic systems into spin chain models} \label{app:mapping}

In this appendix we sketch the main steps of the 
Jordan-Wigner Transformation (JWT)~\cite{lieb61}
mapping the Hubbard/$t-V$ fermionic Hamiltonians with open 
boundary conditions into spin-$1/2$ ladder/chain models.
In particular, we show that
the bath operators injecting or extracting fermions at the two borders
of the system correspond, in the spin-$1/2$ picture, to operators flipping
the edge spins.

We start from the simpler case of the $t-V$ model,
described by Hamiltonian~\eqref{eq:hamTV}.
We first perform a JWT of fermions into hard-core bosons
$(a^\dagger_j , \, a_j)$, defined by
\beq
   a_j \equiv e^{i \pi \sum_{k<j} n_k} c_j =
   \bigg[ \prod_{k=1}^{j-1} (1- 2 n_k) \bigg] c_j,
\eeq
where $n_j = c_j^\dagger c_j$ is the fermion number operator.
The operators $a_j$ for different sites commute, but they are not ordinary
bosonic operators, since at most one boson is allowed on each site.
One can indeed show that $(a_j^\dagger)^2 \ket{0} = 0$
and, on the same site, $\{ a_j, a_j^\dagger \} = 1$.
Moreover, by using
\beq
   \prod_{k=1}^{j-1} (1-2n_k) \prod_{k'=1}^j (1-2n_{k'}) = 1 - 2n_j
\eeq
[it follows from $(1-2n_k)(1-2n_k) = 1$], and the fact that terms with
different
site index commute, we find $n_j \equiv c^\dagger_j c_j= a^\dagger_j a_j$
and $c^\dagger_j c_{j+1} = c^\dagger_j (1-2n_j) c_{j+1} = a^\dagger_j a_{j+1}$.

Then we have to transform the hard-core bosons into spin-$1/2$ particles,
in a representation that identifies, at each site, the state
$\ket{0} \equiv a \ket{1}$ with $\ket{\downarrow}$, and
$\ket{1} \equiv a^\dagger \ket{0}$ with $\ket{\uparrow}$.
If $\sigma_j^\alpha$ denote the corresponding spin-$1/2$ particles
(which of course obey the standard anticommutation rules
$\{ \sigma_j^-, \sigma_{j'}^+ \} = \delta_{j, j'}$),
the explicit mapping is given by:
\beq
   \left\{ \begin{array}{l} a_j^\dagger = \sigma_j^+, \\ a_j = \sigma_j^-, \\
   2 a_j^\dagger a_j = \sigma^z_j +1  \, .\end{array}
   \right.
   \label{eq:hardcore}
\eeq
This leads quite straightforwardly to the following expressions:
\beq
   \left\{ \begin{array}{l}
   n_j n_{j+1} = a^\dagger_j a_j a^\dagger_{j+1} a_{j+1} \\
   \hspace{1.17cm} = \frac{1}{4} \left( \sigma^z_j \sigma^z_{j+1}
   + \sigma^z_j + \sigma^z_{j+1} + 1 \right),  \vspace{1mm} \\
   c^\dagger_j c_{j+1} + c^\dagger_{j+1} c_j =
   a^\dagger_j a_{j+1} + a^\dagger_{j+1} a_j \\
   \hspace{2.45cm} = \frac{1}{2} ( \sigma^x_j \sigma^x_{j+1}
   + \sigma^y_j \sigma^y_{j+1} ) \, .
   \end{array} \right.
   \label{eq:simplif}
\eeq
Substituting them in Eq.~\eqref{eq:hamTV} with open boundary conditions
we finally get
\barr
   \Ham_S & = & -\frac{t}{2} \sum_{j=1}^{N-1} \left( \sigma^x_j \sigma^x_{j+1}
   + \sigma^y_j \sigma^y_{j+1} \right) + \frac{V}{4} \sum_{j=1}^{N-1}
   \sigma^z_j \sigma^z_{j+1} \nonumber \\
   & & - \frac{V}{4} \bigg[ \sigma^z_1 + \sigma^z_N - (N-1) - 2 \sum_{j=1}^N \sigma^z_j
   \bigg] \, ,
   \label{eq:HeisMap}
\earr
that is the Heisenberg Hamiltonian with $J_x \equiv -t/2$ and $J_z \equiv V/4$,
plus an on-site transverse uniform magnetic field of strength $V/2$ and local
transverse fields at the edge spins of the chain.

The Lindblad operators are mapped into spin operators that flip
the outer spins.
Indeed we have $c^\dagger_1 = \sigma^+_1$ and $c_1 = \sigma^-_1$ on
the left side; 
\begin{equation}
c^\dagger_N= e^{i \pi \mathcal{N}_F^{(1)}} \sigma^+_N 
\label{eq:N_F}
\end{equation}
and $c_N = e^{i \pi \mathcal{N}_F^{(1)}} \sigma^-_N$ on the right side
(where $\mathcal{N}_F^{(k)}$ is the number of fermions in
the leftmost $N-k$ sites of the chain).
Since after action of the operator $\sigma^+_N$ in (\ref{eq:N_F})
the right-most site is always occupied, we can just as well rewrite 
Eq.~(\ref{eq:N_F}) as 
\begin{equation}
c^\dagger_N=-e^{i \pi \mathcal{N}_F^{(0)}} \sigma^+_N,
\end{equation}
where $\mathcal{N}_F^{(0)}$ is the total number of spins up 
(occupied fermion states) in the entire lattice.

Now, the crucial observation is that $e^{i\pi \mathcal{N}_F^{(0)}}$ 
is an operator which commutes (anticommutes)
with the fermionic algebra consisting even (odd)
number of fermionic operators.
Since all the terms in our Lindblad equations conserve the parity of
density operators~\cite{prosen08}, i.e. they map, in the Liouville space 
sense, the products of even/odd number of fermionic operators 
$c_j,c^\dagger_j$ into products of even/odd number of such operators,
the Lindblad equation~\eqref{eq:MasterEq} can be restricted
to even-parity density operator subspace only. 
Thus, the factor $e^{i\pi \mathcal{N}_F^{(0)}}$ 
cancels from the Lindblad equation, even though the number
of particles (or magnetization, in the spin language) is not conserved
in an open system. Odd-parity density operator subspace, which
would include terms like $\ket{\rm odd}\bra{\rm even}$, etc.,
which change the number of fermions by an even number,
could be studied with a similar (but not identical) Lindblad equation
with an additional minus sign in all the Lindblad terms.
However, since all the physical observables in question,
say $A=j_k, n_k, \ldots$, are represented as products of even number
of fermionic operators, only the even-parity component
of the density operator affects the expectation values $\ave{A} = \tr [\rho A]$. 
Thus we can safely forget the phase operator $e^{i\pi \mathcal{N}_F^{(0)}}$,
and map Lindblad operators from fermionic language to spin language 
in the Lindblad master equation~\eqref{eq:MasterEq} as: 
$c^\dagger_j \to \sigma^+_j$ and $c_j \to \sigma^-_j$,
keeping in mind the issue of operator-space-parity in case
expectation values of odd-parity operators would be needed.

\begin{figure}[!t]
  \begin{center}
    \includegraphics[scale=0.33]{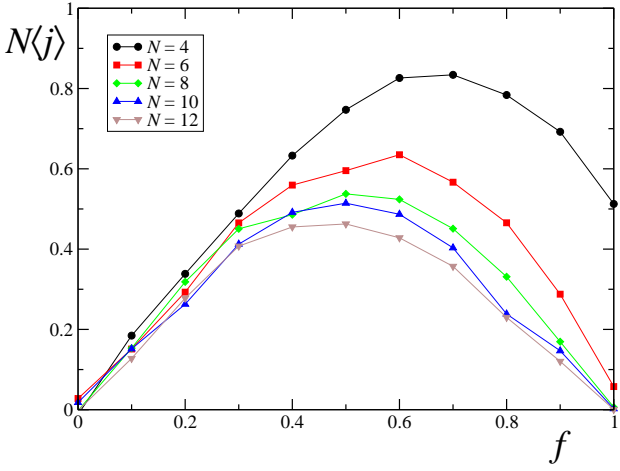}
    \caption{(Color online). Spin current as a function of the driving strength
      for the Hamiltonian in Eq.~\eqref{eq:HeisMap} with $t=2$, $V=8$
      (that corresponds to $J_x=1$, $J_z = 2$, so that $\Delta = 2$).
      In contrast to the standard XXZ model, here
      we added a uniform transverse magnetic field of intensity $V/2$,
      plus two transverse fields of strength $-V/4$ at the border sites of the
      Heisenberg chain. The system-bath coupling is set equal to $\Gamma = 4$;
      the simulation time (QT approach) is $T = 2.5 \times 10^4$.}
    \label{fig:Border_NDC}
  \end{center}
\end{figure}

The charge current becomes then a spin current, while
$\frac{1}{2} (1+\langle \sigma_j^z \rangle)$ is the charge density on site $j$.
We numerically checked that the addition of two local transverse fields on the
border sites and of a uniform magnetic field does not qualitatively affect
the NDC effect, as shown in Fig.~\ref{fig:Border_NDC}.

We now show that the Hubbard model of Eq.~\eqref{eq:Hubbard}
can be mapped into the Heisenberg spin ladder of Eq.~\eqref{eq:HeisLadder}.
Following the same steps for the $t-V$ model, one first performs a double JWT
of spin-up and spin-down fermions into two species
of hard core bosons $(a^\dagger_j , \, a_j)$ and $(b^\dagger_j , \, b_j)$.
Such transformation is defined by
\barr
   a_j & \equiv & e^{i \pi \sum_{k<j} n_{k, \uparrow}} c_{j, \uparrow} \, , \\
   b_j & \equiv & e^{i \pi \sum_{k<j} n_{k, \downarrow}} c_{j, \downarrow} \; .
\earr
Then the hard core bosons are transformed into two species of
spin-$1/2$ particles ($\sigma_j$ and $\tau_j$), in a representation analogous
to that of Eqs.~\eqref{eq:hardcore}.
Proceeding along the same transformations as before, and using $n_{j, \uparrow}
n_{j, \downarrow} =
\frac{1}{2} \left(\sigma^z_j +1 \right) \frac{1}{2} \left(\tau^z_j +1 \right)$,
we finally arrive at the spin ladder Hamiltonian~\eqref{eq:HeisLadder}.
In complete analogy with the $t-V$ model, the Lindblad operators of
Eqs.~\eqref{eq:Lindblad},~\eqref{eq:LindbladRight} are mapped into spin
operators flipping the outer spins of the ladder.

\section{Spin-spin correlations} \label{app:spinspin}

\begin{figure}[!h]
  \begin{center}
    \includegraphics[scale=0.65]{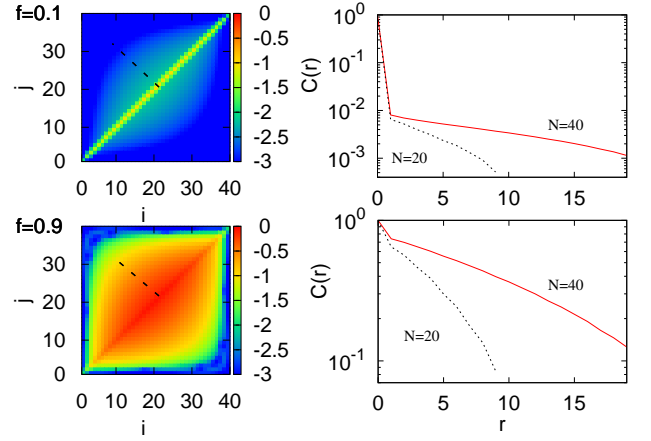}
    \caption{(Color online). Spin-spin correlation function $C(i,j)$ in a chain
      with $40$ spins, for $f = 0.1$ (top row) and $f = 0.9$ (bottom row);
      here we set $\Delta = 2$ and $\Gamma = 4$.
      The code in the left panels denotes $\log_{10} [C(i,j)]$.
      The right plots display correlation function $C(r \equiv \vert i-j \vert)$
      along the diagonal denoted by a dashed line in the left plots; the case
      with $N = 20$ spins is also shown. Data are obtained using the MPO ansatz.}
    \label{fig:Correl}
  \end{center}
\end{figure}

We provide here some numerical data on the behavior of the
steady-state spin-spin correlation functions for the Heisenberg chain
driven far from equilibrium, aimed at investigating the emergence,
in the gapped regime and at strong driving strength $f$,
of a long-range correlation order~\cite{prosen08b}.
The steady-state spin-spin correlation function is defined as:
\beq
  C(i,j) = 
\langle \sigma_i^z \sigma_j^z \rangle-
\langle \sigma_i^z  \rangle
\langle \sigma_j^z  \rangle,
  \label{eq:corel}
\eeq
where $\langle \sigma_i^z \sigma_j^z \rangle = 
\tr{(\sigma_i^z \sigma_j^z \rho_s)}$
and 
$\langle \sigma_i^z  \rangle = \tr{(\sigma_i^z \rho_s)}$,
while averages are taken on the steady state.
Results are plotted in Fig.~\ref{fig:Correl}.
With increasing $f$ and above $f^*$, a dramatic slowing down of the
correlation decay is apparent. 
Unfortunately, a more quantitative understanding of a possible critical
behavior in the large-$f$ regime is at present out of our capabilities,
since it would require the analysis of much bigger systems,
and this is numerically not accessible for the model under consideration.

\section{Solution of the one-magnon model} \label{app:magloc}

The matrix associated with the Hamiltonian~\eqref{eq:Heisenberg}
in the one-magnon basis $\{\ket{1}, \ket{2}, ..., \ket{N} \}$ 
is tridiagonal and reads as follows:
\begin{equation}
H=\left[
\begin{array}{ccccccccc}
 \alpha'   & \beta    & 0        &       &     & ... &       &        &     0   \\
 \beta     & \alpha   & \beta    & 0     &     & ... &       &        &     0   \\
   0       &   \beta  &  \alpha  & \beta &  0  & ... &       &        &     0   \\ 
   \vdots  &          &          &       &     &     &       &        & \vdots  \\ 
   0       &          &          &       &     & ... & \beta & \alpha & \beta   \\
   0       &          &          &       &     & ... & 0     & \beta  & \alpha' 
\end{array}
\right],
\end{equation}
where $\alpha'\equiv (N-3)J_z$, $\alpha\equiv (N-5)J_z$, 
and $\beta\equiv 2J_x$.
Note that the overall magnetization is conserved by the XXZ Hamiltonian,
so the states corresponding to the other spin sectors are not coupled (by the 
autonomous evolution) to the one-magnon sector. 
The eigenvalues of $H$ are given by the roots of the 
characteristic polynomial $D_N(E)\equiv {\rm det}(E\openone - H)$. 

First, it is convenient to solve the eigenvalue problem for $\alpha=\alpha'$.
We define $D_N^{(0)}\equiv D_N \vert_{\alpha=\alpha'}$.
The following recurrence relation holds:
\begin{equation}
   D_j^{(0)}=(E-\alpha) D_{j-1}^{(0)} - \beta^2 D_{j-2}^{(0)}. 
\end{equation}
The general solution to this difference equation is 
\begin{equation}
   D_m^{(0)}=\beta^m \left( A e^{im\phi} + B e^{-im\phi} \right),
\end{equation}
with 
\begin{equation}
  \cos \phi \equiv \left( \frac{E-\alpha}{2\beta} \right)
\end{equation}
and the constants $A$ and $B$ determined from the conditions
\barr
   \displaystyle &D_1^{(0)}=E-\alpha,\nonumber \\
   \displaystyle & D_2^{(0)}=(E-\alpha) D_1^{(0)}-\beta^2. 
\earr
We finally obtain
\begin{equation}
  D_N^{(0)}=\frac{\beta^N \sin [(N+1)\phi]}{\sin\phi},
\end{equation}
whose solutions are
\begin{equation}
  E_m^{(0)}=\alpha+2\beta\cos\left(\frac{m\pi}{N+1}\right),\quad  m=1,...,N.
\end{equation}
These eigenvalues are located in the energy band 
$\alpha - 2 |\beta| < E < \alpha + 2 |\beta|$.
Therefore, $\phi$ is always real and the eigenstates (``Bloch orbitals'') 
\begin{equation}
  |\psi_m^{(0)}\rangle = \sqrt{\frac{2}{N+1}} \sum_{k=1}^N
  \sin\left( \frac{\pi m k}{N+1}\right) \ket{k} , 
\end{equation}
corresponding to the eigenvalues $E_m^{(0)}$ 
are delocalized along the spin chain.

In the case $\alpha\ne \alpha'$ we obtain 
\barr
   D_N & = & D_N^{(0)} +2(\alpha-\alpha') D_{N-1}^{(0)} 
   +(\alpha-\alpha')^2 D_{N-2}^{(0)}  \nonumber \\
   & = & \displaystyle \frac{\beta^N}{\sin\phi} 
   \left\{\sin[(N+1)\phi] \right.               \\
   & & \hspace{0.8cm} - \left. 2 \Delta \sin(N\phi) +
   \Delta^2 \sin[(N-1)\phi]\right\},  \nonumber
\earr
where we have used 
\begin{equation}
  \frac{\alpha' -\alpha}{\beta} = \frac{J_z}{J_x} = \Delta \, .
\end{equation}
The equation $D_N=0$ can be analytically solved for large $N$. 
There always exist at least $N-2$ delocalized solutions lying in the 
energy band between $\alpha-2|\beta|$ and $\alpha + 2 |\beta|$.
The ``molecular orbitals'' 
$\ket{\psi_\pm} \approx \frac{1}{\sqrt{2}}(\ket{\psi_L} \pm \ket{\psi_R})$ 
appear when $|\Delta|>1$.

If $\Delta>1$, we have $\phi=i\chi$ ($\chi\in \mathbb{R})$, 
$e^\chi\approx \Delta$,
\barr
   & \displaystyle \ket{\psi_L} \approx
   \sqrt{\frac{1-e^{-2\chi}}{1-e^{-2N\chi}}}
   \sum_{m=1}^N e^{-(m-1)\chi} \ket{m} \, ,
   \nonumber\\
   & \displaystyle \ket{\psi_R} \approx
   \sqrt{\frac{1-e^{-2\chi}}{1-e^{-2N\chi}}}
   \sum_{m=1}^N e^{-(N-m)\chi} \ket{m} \, .
\earr
The states $\ket{\psi_{L,R}}$ are centered at sites $1$ and $N$,
respectively, and their
localization length $\ell\approx 1/\chi=1/\ln(\Delta)$.
The corresponding eigenvalues are given by  
\begin{equation}
  E_1 \approx E_N \approx \alpha + \beta\left(\Delta+\frac{1}{\Delta}\right).
  \label{eq:loceig}
\end{equation}

\begin{figure}[!t]
  \begin{center}
    \includegraphics[scale=0.31]{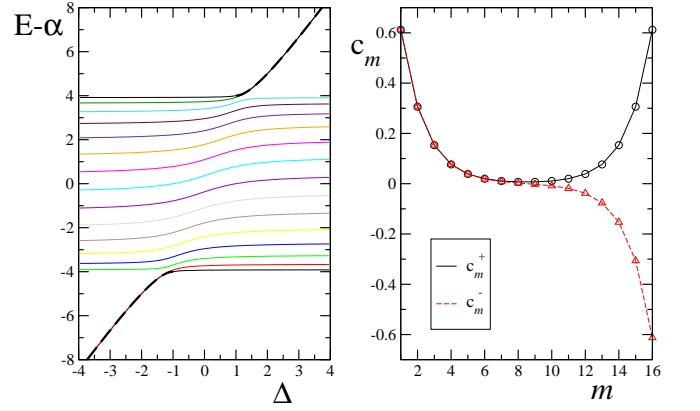}
    \caption{(Color online). The one-magnon model for $N=16$ spins.
      Left panel: energy spectrum as a function of $\Delta$;
      the large-$N$ analytic result~\eqref{eq:loceig} is shown
      by thick dashed lines.
      Right panel: molecular states $\ket{\psi_\pm} = \sum_{m=1}^N (c_m)_\pm \ket{m}$, 
      at $\Delta=2$; the numerically computed coefficients 
      $(c_m)_+$ and $(c_m)_-$ are shown as circles and triangles, 
      the large-$N$ analytic results as full and dashed curves.}
    \label{fig:OneMagnon}
  \end{center}
\end{figure}

If $\Delta<-1$, $\phi=i\chi+\pi$ ($\chi\in \mathbb{R})$, 
$e^\chi\approx -\Delta$,
\barr
   & \displaystyle \ket{\psi_L} \approx
   \sqrt{\frac{1-e^{-2\chi}}{1-e^{-2N\chi}}}
   \sum_{m=1}^N (-1)^m e^{-(m-1)\chi} \ket{m} \, ,
   \nonumber\\
   & \displaystyle \ket{\psi_R} \approx
   \sqrt{\frac{1-e^{-2\chi}}{1-e^{-2N\chi}}}
   \sum_{m=1}^N (-1)^m e^{-(N-m)\chi} \ket{m} \, .
\earr
These states have localization length $\ell\approx 1/\chi=1/\ln(-\Delta)$.
The corresponding eigenvalues are again given by Eq.~\eqref{eq:loceig}.

The gap $\delta_E$ between the energy levels $E_1$ and $E_N$ shrinks
exponentially with the system size:
\begin{equation}
  \delta_E \propto |\langle \psi_L | \psi_R \rangle| \approx 
  \exp\left(-\frac{N}{\ell}\right) \, ,
\end{equation}
so that the coherent tunneling between sites $1$ and $N$ requires 
a time scale which grows exponentially with $N$. 
Therefore, for the purposes of our present investigation
we can say that a spin-flip excitation created at one boundary of a 
ferromagnetic domain remains in practice exponentially localized
over a length $\ell=1/\ln \vert \Delta \vert$.

Numerical illustrations, for a chain of $N=16$ spins, 
of the appearance, for $\Delta|>1$, of two states outside the 
energy band $(\alpha-2|\beta|,\alpha+2|\beta|)$ and of the molecular 
orbitals $\ket{\psi_\pm}$, are provided in Fig.~\ref{fig:OneMagnon}.

Finally, we note that the one-magnon localization 
model discussed in this section has deep similarities 
with a tight-binding model discussed in the context of surface 
physics~\cite{brivio}.

\end{document}